\documentclass[useAMS,usenatbib]{mn2e}

\usepackage{times}  
\usepackage{listings}
\usepackage{graphics}
\usepackage{subfigure}
\usepackage{graphicx}
\usepackage{amssymb}
\usepackage{hyperref}
\usepackage{aas_macros}

\def\simprop{ \lower .75ex \hbox{$\sim$} \llap{\raise .27ex \hbox{$\propto$}} }

\title[Evolution of the gas content]{
Cosmic evolution of the atomic and molecular gas content of galaxies and scaling relations}
\author[Claudia del P. Lagos]{Claudia del P. Lagos, et al. \\
Institute for Computational Cosmology, Department of Physics, University of Durham, South Road, Durham DH1 3LE,\\
United Kingdom}

\title[Evolution of the HI and H$_2$ gas content]{
Cosmic evolution of the atomic and molecular gas content of galaxies}
\author[Claudia del P. Lagos]{
\parbox[t]{\textwidth}{
\vspace{-1.0cm}
Claudia del P. Lagos$^{1}$,
Carlton M. Baugh$^{1}$,
Cedric G. Lacey$^{1}$,
Andrew J. Benson$^{2}$,
Han-Seek Kim$^{3}$,
Chris Power$^{4,5}$
}
\vspace*{6pt} \\
$^{1}$Institute for Computational Cosmology, Department of Physics,
University of Durham, South Road, Durham, DH1 3LE, UK. \\
$^{2}$California Institute of Technology, Pasadena, CA 91125, USA.\\
$^{3}$School of Physics, University of Melbourne, Parkville, VIC 3010, Australia.\\
$^{4}$Department of Physics \& Astronomy, University of Leicester, Leicester LE1 7RH, UK\\
$^{5}$Astronomy and Astrophysics department, University of Western Australia, 35 Stirling Highway, Crawley WA 6009, Australia.
\vspace*{-0.5cm}}

\begin{document}


\pagerange{\pageref{firstpage}--\pageref{lastpage}} \pubyear{2011}

\maketitle

\label{firstpage}

\begin{abstract}
We study the evolution of the cold gas content of galaxies 
by splitting the interstellar medium into 
its atomic and molecular hydrogen components, using the galaxy formation model
{\texttt{GALFORM}} in the $\Lambda$CDM framework. We calculate the 
molecular-to-atomic hydrogen mass ratio, H$_2$/HI, 
in each galaxy using two different approaches; 
the pressure-based empirical relation of Blitz \& Rosolowsky 
and the theoretical model of Krumholz, McKeee \& Tumlinson, and apply them to consistently 
calculate the star formation rates of galaxies. We find that the 
model based on the Blitz \& Rosolowsky law 
predicts an HI mass function, $^{12}\rm CO(1-0)$ luminosity function, 
correlations between the H$_2$/HI ratio and 
stellar and cold gas mass, and infrared-CO  
luminosity relation in good agreement 
with local and high redshift observations. The $\rm HI$ mass function evolves weakly with 
redshift, with the number density of high mass galaxies decreasing with increasing redshift. In the 
case of the H$_2$ mass function, the number density of massive 
galaxies increases strongly from $z=0$ to $z=2$, followed by weak evolution up to $z=4$.
We also find that the $\rm H_2/HI$ ratio of galaxies is strongly 
dependent on stellar and cold gas mass, and also on redshift. 
The slopes of the  
correlations between H$_2$/HI and stellar and cold gas mass 
hardly evolve, 
but the normalisation increases by up to two orders of magnitude from $z=0$ to $z=8$.
The strong evolution in the $\rm H_2$ mass function and the $\rm H_2/HI$ ratio is primarily due to 
the evolution in the sizes of galaxies and secondarily, 
in the gas fractions. The predicted cosmic density evolution of HI agrees with the observed 
evolution inferred from damped-Ly$\alpha$ systems, and is always dominated by the HI content 
of low and intermediate mass halos. We find that previous theoretical studies have largely 
overestimated the redshift evolution of the global $\rm H_2/HI$ ratio  
due to limited resolution. We predict 
a maximum of $\rho_{\rm H_2}/\rho_{\rm HI}\approx 1.2$ 
at $z\approx 3.5$.
\end{abstract}

\begin{keywords}
galaxies: formation - galaxies : evolution - galaxies: ISM - stars: formation
\end{keywords}

\section{Introduction}

Star formation (SF) is a key process in galaxy formation and evolution. 
A proper understanding of SF and the mechanisms regulating it are necessary 
to reliably 
predict galaxy evolution. In recent years there has been a growing interest in modeling 
SF with sub-grid physics in cosmological 
scenarios, in which an accurate description of the
interstellar medium (ISM) of galaxies is needed 
(e.g. \citealt{Springel03}; \citealt{Schaye04}; \citealt{Dutton09}; \citealt{Narayanan09}; \citealt{Schaye10}; \citealt{Cook10}; 
\citealt{Lagos10}; \citealt{Fu10}).  

It has been shown observationally that SF takes place in molecular clouds 
(see \citealt{Solomon05} for a review).
Moreover, the surface density of the star formation rate (SFR) correlates 
with the surface density of molecular hydrogen, H$_2$, in an approximately linear 
fashion, $\Sigma_{\rm SFR}\propto \Sigma_{\rm H_2}$  
(e.g. \citealt{Bigiel08}; \citealt{Schruba10}).
On the other hand, the correlation between $\Sigma_{\rm SFR}$ and the 
surface density of atomic hydrogen, HI, is much weaker. 
Low SFRs have been measured in 
low stellar mass, HI-dominated dwarf and low surface brightness galaxies (e.g. 
\citealt{Roychowdhury09}; \citealt{Wyder09}), and larger SFRs in 
more massive galaxies with more 
abundant H$_2$, such as normal spiral and starburst galaxies 
(e.g. \citealt{Kennicutt98}; \citealt{Wong02}). 

Observational constraints on the 
HI and H$_2$ content of galaxies are now available 
for increasingly large samples. For HI, 
accurate measurements of the $21$~cm emission in large 
surveys of local galaxies have been presented by \citet{Zwaan05} using the HI Parkes 
All-Sky Survey (HIPASS; \citealt{Meyer04}) and more recently by \citet{Martin10} 
using the Arecibo Legacy Fast ALFA Survey (ALFALFA; \citealt{Giovanelli05}). From these surveys 
it has been possible to probe the HI mass function (MF) down to 
HI masses of $M_{\rm HI}\approx 10^{6} M_{\odot}$ and to estimate 
the global HI mass density at $z=0$, 
$\Omega_{\rm HI}=3.6-4.2 \times 10^{-4}$ in units of the present day
critical density.
These HI selected galaxies are also characterised by  
weaker clustering (e.g. \citealt{Basilakos07}; \citealt{Meyer07}) 
than optically selected samples (e.g. \citealt{Norberg01}), indicating that local HI 
selected galaxies 
are preferentially found in lower mass haloes than their optical counterparts.
At high redshift information about HI 
is very limited since it is based mostly on absorption-line measurements 
in the spectra of quasi-stellar object (QSO;  
e.g. \citealt{Peroux03}; \citealt{Prochaska05}; 
\citealt{Rao06}; \citealt{Guimaraes09}; \citealt{Noterdaeme09}; see \citealt{Rauch98} 
for a review of this technique). These observations suggest very 
little evolution of $\Omega_{\rm HI}$ up to $z\approx 5$. 
Intensity mapping of the $21$~cm emission line 
is one of the most promising techniques to 
estimate HI mass abundances at high redshifts. 
This technique has recently been applied 
to estimate global HI densities 
at intermediate redshifts ($z\lesssim 0.8$), 
and has given estimates in agreement with the ones inferred 
from absorption-line measurements (e.g.  \citealt{Verheijen07}; 
\citealt{Lah07}, 2009; \citealt{Chang10}). 

To study H$_2$, 
it is generally necessary to use emission from other molecules as tracers, 
since H$_2$ 
lacks a dipole moment, making emission from this 
molecule extremely weak and hard 
to detect in interstellar gas, which is typically cold. 
The most commonly used proxy for H$_2$ is the $^{12}\rm CO$ molecule (hereafter `CO'), which is the second most abundant 
molecule in the Universe. \citet{Keres03} reported the first attempt to derive the 
local luminosity function (LF) of 
$\rm CO(1-0)$ (the lowest energy transition of the $\rm CO$ molecule) 
from which they inferred the H$_2$ MF and the local 
$\Omega_{\rm H_2}=1.1\pm 0.4\times 10^{-4} \, h^{-1}$, assuming a 
constant $\rm CO(1-0)$-H$_2$ conversion factor. 
It has not yet been possible 
to estimate the cosmic H$_2$ abundance at high redshift. However, a 
few detections of H$_2$ absorption in the lines-of-sight to QSOs have been reported 
(e.g. \citealt{Noterdaeme08}; \citealt{Tumlinson10}; \citealt{Srianand10}), as
well as $\rm CO$ detections in a large number of luminous star-forming galaxies (e.g.
\citealt{Greve05}; \citealt{Geach09}; \citealt{Daddi10}; \citealt{Tacconi10}). 

Measurements of the HI and H$_2$ 
mass content, as well as other galaxy 
properties, are available in relatively large 
samples of local galaxies (running into a few hundreds), allowing 
the characterisation of scaling relations between 
the cold gas and the stellar mass content.
From these samples it has been possible to determine 
that the molecular-to-atomic gas ratio 
correlates with stellar mass, and 
that there is an anti-correlation between the HI-to-stellar mass ratio and stellar mass 
(e.g. \citealt{Bothwell09}; \citealt{Catinella10}; \citealt{Saintonge11}). 
However, these correlations exhibit large scatter 
and are either subject to biases in the construction 
of observational samples, such as 
inhomogeneity in the selection criteria, or sample 
a very narrow range of galaxy properties.

The observational constraints on HI and H$_2$ at higher redshifts 
will improve dramatically over the next decade with the 
next generation of radio and sub-millimeter telescopes such as the Australian SKA 
Pathfinder (ASKAP; \citealt{Johnston08}), the Karoo Array Telescope
(MeerKAT; \citealt{Booth09}) and the Square Kilometre Array 
(SKA; \citealt{Schilizzi08}) which aim to detect $21$ cm emission from HI,  
and the Atacama Large Millimeter Array 
(ALMA; \citealt{Wootten09}) and the Large Millimeter Telescope (LMT; \citealt{Hughes10})  
which are designed to detect emission from molecules. 
Here we investigate the predictions of galaxy formation models for the evolution of the HI and H$_2$ 
gas content of galaxies, taking advantage of the development of realistic SF models 
(e.g. \citealt{MacLow04}; \citealt{Pelupessy06}; \citealt{Blitz06}; \citealt{Krumholz09}; \citealt{Pelupessy09}; 
see \citealt{McKee07} for a review). In our approach, a successful model 
is one that, at the same time, reproduces the observed stellar masses, luminosities, morphologies 
and the atomic and molecular 
 gas contents of galaxies at the present day. Using such a  
model, it is reasonable 
to extend the predictions to follow the evolution 
of the gas contents of galaxies 
towards high redshift. 

{ Until recently, the ISM of galaxies 
in semi-analytic models was treated as a single star-forming phase 
(e.g. \citealt{Cole00}; \citealt{Springel01}; \citealt{Cattaneo08}; \citealt{Lagos08};
 \citealt{Somerville08}). The first attempts to predict the separate HI and H$_2$ contents 
of galaxies in semi-analytic models postprocessed the output of 
single phase ISM treatments  
to add this information {\it a posteriori} (e.g. \citealt{Obreschkow09}; \citealt{Power10}). It 
was only very recently that a proper fully self-consistent treatment of the ISM and SF in galaxies 
throughout the cosmological calculation was made (\citealt{Cook10}; \citealt{Fu10}; 
\citealt{Lagos10}).}
We show 
{ in this work} that this consistent treatment 
is necessary to 
make progress in understanding the gas contents of galaxies.

We use the semi-analytical model {\texttt{GALFORM}} (\citealt{Cole00}) 
in a $\Lambda$ cold dark matter ($\Lambda$CDM) 
cosmology with the improved treatment of SF implemented by \citet{Lagos10}, 
which explicitly splits the hydrogen content in the ISM 
of galaxies into HI and H$_2$.
Our aims are (i) to study whether the models are able 
to predict HI and H$_2$ MFs in agreement 
with the observed ones at $z=0$, (ii) to follow 
the evolution of the MFs towards high-redshift, 
(iii) to compare with the observational 
results described above and (iv) to study 
scaling relations of H$_2$/HI with 
galaxy properties. By doing so, it is possible to establish 
which physical processes are responsible 
for the evolution of HI and H$_2$ in galaxies. 

This paper is organized as follows. In $\S$\ref{modelssec} we describe the main characteristics 
of the {\texttt{GALFORM}} model.
In $\S$\ref{localU} we present local universe scaling relations and compare with 
available observations. In $\S$\ref{MFs} we present the local HI and H$_2$ MFs, 
and the Infrared (IR)-CO luminosity relation.   
We also investigate which galaxies dominate the HI and H$_2$ MFs and predict their   
evolution up to $z=8$, and study the HI mass density 
measured in stacked samples of galaxies.
In $\S$\ref{ScalingRels}, we present predictions 
for the scaling relations of the H$_2$/HI ratio with galaxy properties and
analyse the mechanisms which underly these relations.
$\S$\ref{cosmicevo} presents 
the evolution of the cosmic densities of HI and H$_2$, and 
compares with observations. We also determine which range of halo mass dominates 
these densities.
Finally, we discuss our results and present our 
conclusions in $\S$\ref{conclusion}.

\section{Modelling the two-phase cold gas in galaxies}\label{modelssec}

We study the evolution of the cold gas content of galaxies by splitting the
ISM into atomic and molecular hydrogen components in the {\texttt{GALFORM}}
semi-analytical model of galaxy formation (\citealt{Cole00}; 
\citealt{Benson03}; \citealt{Baugh05};
\citealt{Bower06}; \citealt{Benson10}; \citealt{Lagos10}). 

The model 
takes into account the main physical processes
that shape the formation and evolution of galaxies. These are: (i) the collapse
and
merging of dark matter (DM) halos, (ii) the shock-heating and radiative cooling
of gas inside
DM halos, leading to the formation of galactic disks, (iii) quiescent star
formation in galaxy disks, (iv) feedback
from supernovae (SNe), from active galactic nucleus (AGN) 
heating and from photo-ionization of the
inter-galactic medium (IGM), (v) chemical 
enrichment of stars and gas, and (vi) galaxy mergers driven by
dynamical friction within common DM halos which can trigger bursts of SF,
and lead to the formation of spheroids (for a review of these
ingredients see \citealt{Baugh06}; \citealt{Benson10b}). 
Galaxy luminosities are computed from the predicted star formation and
chemical enrichment histories using a stellar population synthesis model.
Dust extinction at different wavelengths is calculated
self-consistently from the gas and metal contents of
each galaxy and the predicted scale lengths of the disk and bulge components
using a radiative transfer model 
(see \citealt{Cole00} and \citealt{Lacey11}). 

We base our study on the
scheme of Lagos et al. (2011; hereafter L11)
in which parameter-free SF laws were implemented in {\texttt{GALFORM}}.
In the next four subsections
we briefly describe
the dark matter (DM) merger trees, the {\texttt{GALFORM}} models 
considered, the main features introduced by L11 into
{\texttt{GALFORM}}, and the importance of the 
treatment of the ISM made in this work. 

\subsection{Dark matter halo merger trees}

{\texttt{GALFORM}} requires 
the formation histories of DM halos to model galaxy formation (see \citealt{Cole00}). 
To generate these histories we use an improved version of the Monte-Carlo 
scheme of \citet{Cole00}, which was derived by \citet{Parkinson08}. 
The Parkinson et al. scheme is tuned to match merger trees extracted from the 
Millennium simulation of \citet{Springel05}.
In this approach, realizations of the merger histories 
of halos are generated over a range of halo masses. 
The range of masses simulated changes 
with redshift in order to follow a 
representative sample of halos, 
which cover a similar range of abundances at each epoch.

By using Monte-Carlo generated merger histories, we can extend the range of halo masses considered beyond the 
resolution limit of the
Millennium simulation, in which 
the smallest resolved halo mass is $\approx 10^{10}
h^{-1} \rm M_{\odot}$ at all redshifts. 
This is necessary to make an accurate census of the global cold gas density of the universe which is dominated 
by low mass galaxies in low mass haloes (\citealt{Kim10}; \citealt{Power10}; L11).

We adopt a minimum halo mass of $M_{\rm halo}=5 \times 10^{8} h^{-1} M_{\odot}$ at
$z=0$ 
to enable us to predict cold gas mass structures down to the current observed
limits (i.e. $M_{\rm HI}\approx 10^{6} M_{\odot}$, \citealt{Martin10}). At
higher redshifts, { this lower limit is scaled with redshift 
to roughly track the evolution of the break in the halo mass function, so that 
we simulate objects with a comparable range of space densities at each redshift. This 
allows us to follow a representative sample of dark matter halos},  
ensuring that we resolve the structures rich in cold
gas at every redshift. \citet{Power10} showed that
using a fixed tree resolution, as imposed by $N$-body simulations, can lead to 
a substantial underestimate of the gas 
content of the universe at $z\gtrsim 3$. 

{ The cosmological parameters are input parameters for the 
galaxy formation model and influence the parameter values adopted to describe  
the galaxy formation physics. The two models used in this paper have somewhat 
different cosmological parameters for historical reasons; 
these cannot be homogenized without revisiting the choice 
of the galaxy formation parameters. 
The parameters used in \citet{Baugh05} 
are a present-day 
matter density of $\Omega_{\rm m}=0.3$, a cosmological constant $\Omega_{\Lambda}=0.7$, 
a baryon density of $\Omega_{\rm baryons}=0.04$, a Hubble constant $h=0.7$ and 
a power spectrum normalization of $\sigma_8=0.93$. In the case of 
\citet{Bower06}, $\Omega_{\rm m}=0.25$, $\Omega_{\Lambda}=0.75$, $\Omega_{\rm baryons}=0.045$, $h=0.73$ and 
$\sigma_8=0.9$.}

\subsection{Galaxy formation models}

We use as starting points the 
\citet{Baugh05} and \citet{Bower06} models, hereafter 
referred to as Bau05 and Bow06 respectively.
The most important differences between these models reside in the mechanism used to suppress SF in
massive galaxies and the form of the stellar initial mass function (IMF) adopted in starbursts. 
The Bau05 model invokes superwinds driven by SNe, 
which expel gas from the halo with a mass ejection rate 
proportional to the SFR. The Bow06 model includes a treatment of the
heating of halo gas by AGN, 
that suppresses gas cooling in haloes where the central black
hole (BH) has an Eddington luminosity which exceeds the cooling luminosity by a
specified factor. The Bau05 model 
assumes that the IMF is top-heavy during starbursts 
(driven by galaxy mergers) 
and has a \citet{Kennicutt83} form during quiescent SF in galactic disks. The Bow06 model assumes 
a universal
Kennicutt IMF (see \citealt{Almeida07}, \citealt{Gonzalez09} and
\citealt{Lacey11} for a detailed
comparison between the two models). Both models give good agreement with the observed $b_J$- and K-band 
LFs at $z=0$, but only the Bow06 model agrees with the K-band 
LF at $z>0.5$, regardless of the SF law used (see L11). 
The insensitivity of the LF of galaxies in optical and near-infrared 
(IR) bands to the choice of the SF law
was already noted by L11 and interpreted as the result of an interplay between the different
channels of SF activity in the model (i.e. burst and quiescent modes).

An important modification made in this work with respect 
to the original assumptions of the Bau05 and Bow06 models resides in the
parameters for the reionisation model, following \citet{Lacey11}.
It is assumed 
that no gas is allowed to cool in haloes with a
circular velocity below $V_{\rm crit}$ at redshifts below $z_{\rm
reion}$ \citep{Benson03}. 
Taking into account recent simulations by \citet{Okamoto08} and observational 
constraints on the reionisation redshift \citep{Spergel03}, we adopt 
$V_{\rm crit}=30\,\rm km \,s^{-1}$ and $z_{\rm reion}=10$, in contrast with the values adopted 
by 
\citet{Baugh05} and \citet{Bower06} ($V_{\rm crit}=50-60\,\rm km \,s^{-1}$ and $z_{\rm reion}=6$). 
Even though this affects the gas content of low mass halos, the changes are not significant for the results 
shown in this paper.

\subsection{The interstellar medium and star formation in galaxies}

We use the SF scheme implemented in {\texttt{GALFORM}} by L11. 
L11 tested different SF laws in which the neutral hydrogen in the ISM 
is split into HI and H$_2$. 
Two of the most promising are: (i) the \citet{Blitz06} empirical SF law and (ii) the 
theoretical SF law of \citet{Krumholz09}. We briefly describe these below. 

(i) The empirical SF law of Blitz \& Rosolowsky is of the form,
\begin{equation}
\Sigma_{\rm SFR} = \nu_{\rm SF} \,\rm f_{\rm mol} \, \Sigma_{\rm gas},
\label{Eq.SFR}
\end{equation}
\noindent where $\Sigma_{\rm SFR}$ and $\Sigma_{\rm gas}$ are the surface
densities of SFR and total cold gas mass, respectively, 
$\nu_{\rm SF}$ is the inverse of the SF 
timescale for the molecular gas and $\rm f_{\rm mol}=\Sigma_{\rm mol}/\Sigma_{\rm gas}$ is the
molecular to total gas mass surface density ratio. The molecular and total gas 
contents include the contribution from helium, while HI and H$_2$ only include  
hydrogen (which in total corresponds 
to a fraction $X_{\rm H}=0.74$ of the overall cold gas mass). 
The ratio $\rm f_{\rm mol}$ depends on
the internal hydrostatic pressure as $\Sigma_{\rm H_2}/\Sigma_{\rm HI}=\rm f_{\rm mol}/(f_{\rm mol}-1)=
(P_{\rm
ext}/P_{0})^{\alpha}$. Here $\rm k_{\rm B}$ is Boltzmann's constant  
and the values of $\nu_{\rm SF}=5.25\pm 2.5\times10^{-10}\, \rm
yr^{-1}$ (\citealt{Leroy08}; \citealt{Bigiel11}),
$\alpha=0.92\pm 0.07$ and $\rm log(P_{0}/k_{\rm B} [\rm cm^{-3} K])=4.54\pm0.07$
\citep{Blitz06} are 
derived from fits to the observational
data. { To calculate $P_{\rm
ext}$, we use the approximation from \citet{Elmegreen93},

\begin{equation}
P_{\rm ext}\approx \frac{\pi}{2} G \Sigma_{\rm gas} \left[ \Sigma_{\rm gas} +
\left( \frac{\sigma_{\rm g}}{\sigma_{\star}}\right)\Sigma_{\star}\right], 
\label{Pext}
\end{equation}

\noindent where $\Sigma_{\rm gas}$ and $\Sigma_{\star}$ are
the total surface densities of the gas and stars, respectively, and
$\sigma_{\rm g}$ and $\sigma_{\star}$ are their respective vertical velocity
dispersions. We assume a constant gas velocity
dispersion, $\sigma_{\rm g}=10\, \rm km\,\rm s^{-1}$, following recent 
observational results \citep{Leroy08}. In the 
case of $\sigma_{\star}$, we assume self-gravity of the stellar disk in the vertical direction,
$\sigma_{\star}=\sqrt{\pi \rm G\, \rm h_{\star}\Sigma_{\star}}$, where 
$h_{\star}$ is the disk height (see L11 for details). We estimate $h_{\star}$ 
by assuming that it is proportional to the radial exponential scale length
of the disk, as observed in local spiral galaxies, with $\rm R_{\rm
eff}/h_{\star} \approx 7.3 \pm 1.2$ \citep{Kregel02}.}  
Hereafter, we will refer to the version of the model in which the Blitz \& Rosolowsky
empirical SF law is applied by appending BR to the model name. 

(ii) In the theoretical SF law of \citet{Krumholz09}, 
$\rm f_{\rm mol}$ in Eq.~\ref{Eq.SFR}   
depends on the balance
between the dissociation of molecules due to the far-UV interstellar
radiation, and their formation on the
surfaces of dust grains, and 
$\nu_{\rm SF}$ the 
inverse of the time required to convert all of the gas in a cloud 
into stars.
In their prescription,
$\rm f_{\rm mol}$ depends on the metallicity of the gas and the surface density
of a molecular cloud, $\Sigma_{\rm complex}$. Krumholz et al. relate $\Sigma_{\rm complex}$ to 
the smooth gas surface density, $\Sigma_{\rm gas}$, through a clumping factor
that is set to $c=5$ to reproduce the observed $\Sigma_{\rm SFR}-\Sigma_{\rm gas}$ relation. 
In summary, the Krumholz et al. SF law is,

\begin{equation}
\Sigma_{\rm SFR}= \nu_{\rm SF}(\Sigma_{\rm gas})\, f_{\rm mol}\, \Sigma_{\rm gas}, 
\end{equation}
\noindent where
\begin{eqnarray}
\nu_{\rm SF}(\Sigma_{\rm gas}) & = & 
\nu^{0}_{\rm SF}\times 
\left(\frac{\Sigma_{\rm gas}}{\Sigma_{\rm 0}}\right)^{-0.33} \quad {\rm for } \quad \Sigma_{\rm gas} < \Sigma_{\rm 0} \nonumber \\
& = & \nu^{0}_{\rm SF}\times \left(\frac{\Sigma_{\rm gas}}{\Sigma_{\rm
    0}}\right)^{0.33}  \quad \quad {\rm for} \quad \Sigma_{\rm gas} > \Sigma_{\rm 0} 
\label{sfreq2}
\end{eqnarray}
\noindent Here $\nu^{0}_{\rm SF}=3.8\times10^{-10}\, \rm yr^{-1}$ and 
$\Sigma_{\rm 0}=85\,M_{\odot}\,\rm pc^{-2}$. Hereafter, we will denote the
version of the model where the Krumholz, McKee \& Tumlinson 
theoretical SF law is applied by KMT. 

We split the total hydrogen mass component into its atomic and molecular forms and 
calculate the SFR based on the molecular gas mass. { It is important 
to note that in the BR SF law the inverse of the SF timescale for the molecular 
gas, $\nu_{\rm SF}$ (Eq.~\ref{Eq.SFR}), 
is constant, and therefore the SFR directly  
depends only on the molecular content, and only indirectly on the total cold gas content through 
the disk pressure. However, in the KMT SF law, $\nu_{\rm SF}$ is a function 
of cold gas surface density (Eq.~\ref{sfreq2}), and therefore the SFR does not only 
depend on the H$_2$ content, but also on the cold gas surface density in a non-linear fashion.} 

{ 
For starbursts the situation is less clear. 
Observational uncertainties, such as 
the conversion factor between CO and H$_2$ in starbursts, and the 
intrinsic compactness of star-forming regions, have not allowed
a good characterisation of the SF law (e.g. \citealt{Kennicutt98}; \citealt{Genzel10};
\citealt{Combes11}).
Theoretically, it has been suggested that the SF law in starbursts is different from 
that in normal star-forming galaxies: the
relation between $\Sigma_{\rm H_2}/\Sigma_{\rm HI}$ and gas pressure 
is expected to be dramatically different in environments of very high 
gas densities typical of starbursts 
(\citealt{Pelupessy06}, \citealt{Pelupessy09}), where the ISM is predicted 
to be always dominated by H$_2$ independently of the gas pressure. 
For these reasons we choose to apply the BR and KMT
laws only during quiescent SF (fuelled by cooled
gas accretion into galactic disks) and retain 
the original SF prescription for
starbursts (see \citealt{Cole00} and L11 for details).} 
In the latter, the SF timescale is proportional to the bulge dynamical
timescale above a minimum floor value 
and involves the whole cold gas content of the galaxy, $\rm SFR={\it M}_{\rm cold}/\tau_{\rm SF}$ 
(see \citealt{Granato00} and \citealt{Lacey08} for details). Throughout this work 
we assume that in starbursts, the cold gas content is fully molecular, $\rm f_{\rm mol}=1$. 
Note that this is similar to assuming that the BR pressure-law 
holds in starbursts (except with a different $\nu_{\rm SF}$) 
given that large gas and stellar densities lead to 
 $\rm f_{\rm mol}\approx1$.

\subsubsection{Radial profiles of atomic and molecular hydrogen}
\begin{figure}
\begin{center}
\includegraphics[trim = 3mm 3mm 1.5mm 1.5mm,clip,width=0.5\textwidth]{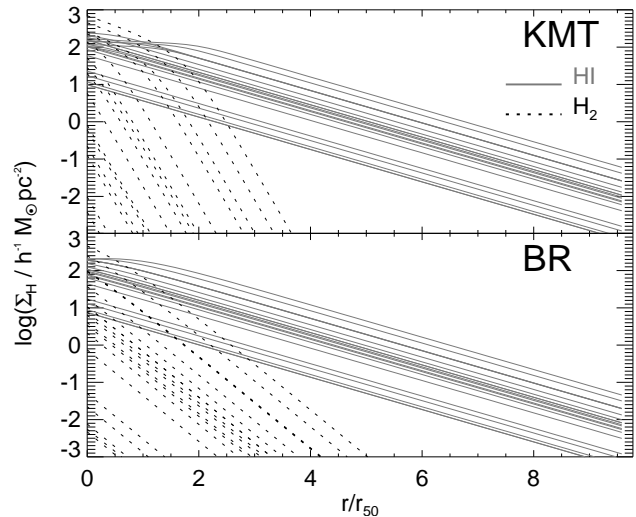}
\caption{Surface density of HI and H$_2$ (solid and dashed lines, respectively) as a 
function of radius in units of the half-mass radius for 
randomly chosen galaxies in the Bow06 model,  
applying the KMT (top panel) and BR (bottom panel) SF laws.} 
\label{HIH2prof}
\end{center}
\end{figure}

In order to visualize the behaviour of the 
HI and H$_2$ components of the ISM of galaxies predicted in the models, we
have taken the output of the original Bow06 
model and postprocessed it to calculate 
the HI and H$_2$ surface density profiles based on the expressions above. Fig.~\ref{HIH2prof} shows 
the surface density profiles of HI and H$_2$ for randomly chosen model spiral galaxies on 
applying the KMT (top panel) or the BR (bottom panel) laws.
The HI extends to much larger radii than the H$_2$.
This is a consequence of the total gas surface density 
dependence of the H$_2/$HI ratio in 
both SF laws. Note that the KMT SF law gives a 
steeper decrease in the radial profile of H$_2$ compared 
to the BR SF law. 
The BR SF law depends on the 
surface density of gas and stars, through the disk pressure, while 
the KMT SF law depends on the surface density 
and metallicity of the gas. 

{ We have compared our predictions for the size of the HI and H$_2$ disks with 
observations. The models predict a correlation between the HI isophotal radius, $l_{\rm HI}$, 
defined at a HI surface density of $1 M_{\odot}\, \rm pc^{-2}$, 
and the enclosed HI mass, $M_{\rm HI}(l_{\rm HI})$, which is in 
very good agreement with observations, irrespective of the SF law applied. Both models predict 
$M_{\rm HI}(l_{\rm HI})/M_{\odot}\approx 2\times 10^7 (l_{\rm HI}/\rm kpc)^{1.9}$, while 
the observed relation for Ursa Major is 
$M_{\rm HI}(l_{\rm HI})/M_{\odot}\approx 1.8\times 10^7 (l_{\rm HI}/\rm kpc)^{1.86}$ (e.g. \citealt{Verheijen01}). 
On the other hand, the median of the relation between the exponential scale length 
of the H$_2$ and HI disks predicted by the model 
is $R_{\rm H_2}\approx 0.4\, R_{\rm HI}$, while 
the relation inferred by combining the observational results of \citet{Regan01} and \citet{Verheijen01} 
is $R_{\rm H_2}=(0.44\pm 0.12)\, R_{\rm HI}$. 
The model agrees well with the observed relations indicating that, in general terms, 
the ISM of modelled galaxies looks realistic.}

\subsection{Consistent calculation or postprocessing?}

Previous attempts to calculate the separate HI and H$_2$ contents  
of galaxies in a cosmological scenario 
have been made by postprocessing the output of existing semi-analytic models using 
specifically the BR SF prescription  
(e.g. \citealt{Obreschkow09}; \citealt{Power10}). 
Here we show that a self-consistent calculation of the ISM of galaxies, { in which 
the new SF laws are included in the model}, is necessary throughout 
in order to explain the observed gas properties of galaxies. To do this 
we choose as an example the Bow06 model and the BR SF law. 

\begin{figure}
\begin{center}
\includegraphics[trim = 5.5mm 5.5mm 2mm 2mm,clip,width=0.49\textwidth]{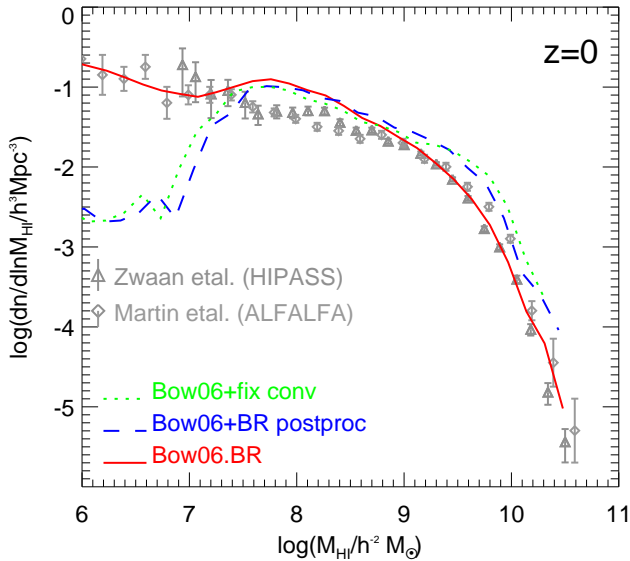}
\caption{The HI mass function at $z=0$ for the original Bow06 model when a constant 
$M_{\rm mol}/M_{\rm cold}=0.27$ ratio is assumed (dotted line), 
when a variable conversion factor based on the BR pressure law is assumed (dotted line) 
and for the Bow06.BR model 
(solid line), where 
the BR SF law is consistently applied throughout the calculation.
Symbols show observational results at $z=0$ from \citet{Zwaan05}
using HIPASS and \citet{Martin10} using ALFALFA, as
labelled.}
\label{HIcompall}
\end{center}
\end{figure}

Fig.~\ref{HIcompall} shows the HI MF in the original Bow06 model when a fixed conversion 
ratio of $M_{\rm mol}/M_{\rm cold}=0.27$ 
is assumed (dotted line; \citealt{Baugh04}), 
when a variable conversion factor is calculated based on the BR pressure law (dashed line) and 
in the Bow06.BR model (solid line), where the BR SF law is applied consistently throughout the calculation. 
Symbols show a compilation of observational data. 
The original Bow06 model with a fixed H$_2$/HI conversion gives a poor match to the observed HI MF. 
Postprocessing the output of this model to implement a variable H$_2$/HI ratio gives a different 
prediction which still disagrees with the observations. If we consistently apply the BR SF law 
throughout the whole galaxy formation model, there is a substantial change in the model prediction 
which is also in much better agreement with the observations. These differences are mainly due to 
the fact that the SFR in the prescription of Bow06 
scales linearly with the total cold gas mass  
(as in starbursts, see $\S 2.3$), while in the case
of the BR SF law this dependence is non-linear. { 
Note that 
a threshold in 
gas surface density below which galaxies are not allowed to form stars, 
as originally proposed by \citet{Kennicutt89} 
and as assumed in several other semi-analytic models (e.g. \citealt{Croton06}; \citealt{Tecce10}), 
produces a much less pronounced, but still present dip (see L11). Thus, the 
linearity of the SFR with cold gas mass is the main driver of the strong dip at low HI masses 
in the original model.
Remarkably, our new modelling helps reproduce the observed number density even down 
to the current limits, $M_{\rm HI}\approx 10^6\, h^{-2}\, M_{\odot}$.} 

The new SF scheme to model the ISM of galaxies represents
a step forward in understanding the gas content of galaxies. In the rest of the paper we
make use of the models where the parameter-free SF laws are applied throughout the full calculation.

\section{Scaling relations for the atomic and molecular contents of galaxies in the local universe}\label{localU}

{ Here we present the model predictions for various scaling relations 
between H$_2$ and HI and other galaxy properties and 
compare with observations. In $\S 3.1$ we show 
how the H$_2$/HI ratio scales with stellar and cold gas mass. In $\S 3.2$ 
we present predictions for this ratio as a function of galaxy morphology, 
and in $\S 3.3$, we study dependence of the HI and H$_2$ masses on stellar mass.}

\subsection{The dependence of H$_2$/HI on galaxy mass}

\begin{figure}
\begin{center}
\includegraphics[width=0.5\textwidth]{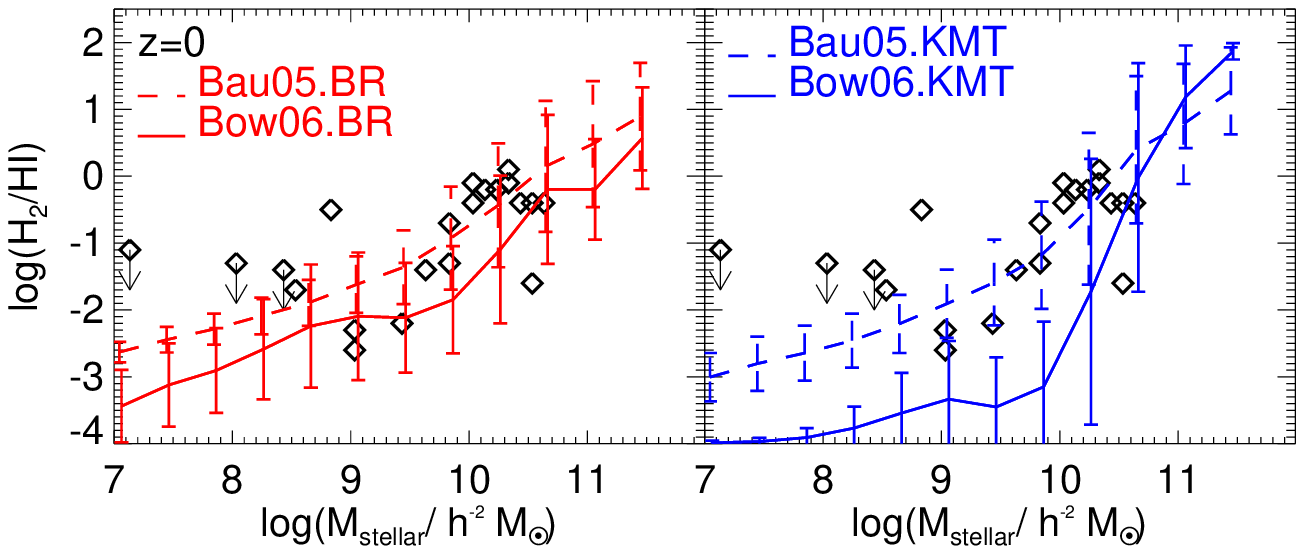}
\includegraphics[width=0.5\textwidth]{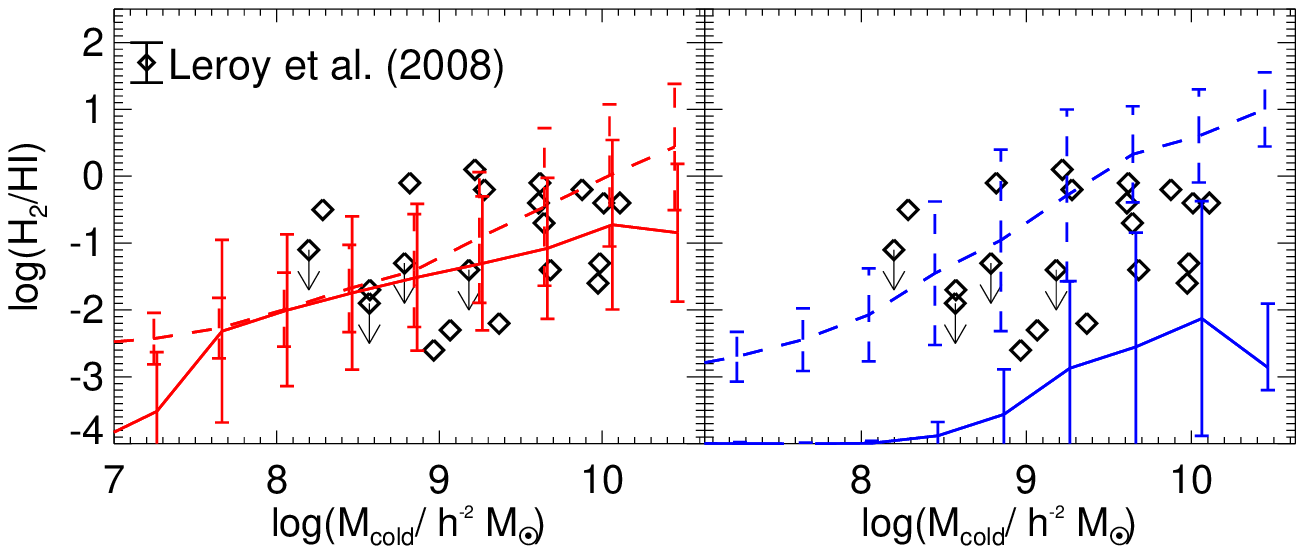}
\caption{Molecular-to-atomic hydrogen ratio, H$_2$/HI, 
as a function of stellar mass (top panels) 
and total cold gas mass ($M_{\rm cold}=\rm He+HI+H_{2}$; bottom panels) 
for the two model variants, the Bow06 (solid lines) and Bau05 (dashed lines) models, when applying the 
BR (left panels) and the KMT (right panel) SF laws. 
Lines show the medians of the predicted distributions, while
errorbars show the $10$\% and $90$\% percentiles. 
Grey symbols show observations of spiral and irregular galaxies from 
\citet{Leroy08}. Symbols with arrows represent upper limits in the observed sample.
Indicative errorbars in the observational sample are shown in the bottom-left panel.}
\label{Scaling1}
\end{center}
\end{figure}

The prescriptions described in $\S 2$ that split the ISM into its 
atomic and molecular components 
enable the model to directly predict correlations between the H$_2$/HI ratio
and galaxy properties. Fig.~\ref{Scaling1} shows the H$_2$/HI ratio as a 
function of stellar mass (top panels) and total cold gas mass (bottom panels) 
at $z=0$ for the Bow06 (solid lines) and the Bau05 (dashed lines) models using the 
BR and the KMT SF laws. The model predictions are compared to local 
observational estimates from \citet{Leroy08}. The errorbars on the model show
the $10$ and $90$ percentiles of the model distribution in different mass bins. In the case of the observations, 
indicative errorbars on the H$_2$/HI estimate due to the uncertainty in the CO-H$_2$ conversion 
factor are shown at the top of the top-left panel 
(i.e. the difference in the value inferred for 
starbursts and the Milky-Way, $\approx 0.6$dex; see $\S 4.2$).{ Note 
that in this and subsequent figures where we compare with observations, 
we plot masses in units of $h^{-2} M_{\odot}$ to match observational units. 
For other plots we use the simulation units, $h^{-1} M_{\odot}$.}
 
Leroy et al. estimated stellar masses from $3.6\mu$m luminosities, 
which were transformed into $K$-band luminosities 
using an empirical conversion. Stellar masses 
were then calculated from the relation of \citet{Bell03b} between the 
stellar mass-to-light ratio in the $K-$band and $B-V$ colour\footnote{Given 
that Leroy et al. adopt a \citet{Kroupa01} IMF, we
apply a small correction of $1.02$ to adapt their stellar masses to account for
our choice of a \citet{Kennicutt83} IMF.}.  
Thus the error on the stellar mass might be as large as a factor of
$1.5$ (considering the dispersion of $0.15$~dex in
the $B-V$ vs. $K$-band relation). 

The Bow06 model with the BR SF law predicts H$_2$/HI ratios 
which are in very good agreement with the observed
ones. The Bow06.KMT model fails to match the observed H$_2$/HI ratios at all stellar
 and cold gas masses. This is due to the sharp decline 
of the radial profile of H$_2$ (see 
Fig.~\ref{HIH2prof}), which results in lower global 
H$_2$/HI ratios in disagreement with the 
observed ones. In the case of the Bow06.BR model, it might appear 
unsurprising that this model is in good
agreement with the observed correlation given that it is based on  
the empirical correlation between H$_2$/HI and the 
hydrostatic pressure in the disk (see $\S 2$). 
However, it is only because we are able to reproduce other galaxy 
properties,
such as stellar mass functions (see $\S 2$), gas fractions 
and, approximately galaxy sizes (L11), that we also predict the observed 
trend in the scaling relations shown in Fig.~\ref{Scaling1}. 

Both the Bau05.BR and Bau05.KMT models also give good agreement 
with the observed H$_2$/HI ratios. The success of the Bau05.KMT
model, in contrast 
with the Bow06.KMT model, is mainly due to the 
higher gas masses 
and metallicites predicted in the former model. 
However, the Bau05.BR and Bau05.KMT models fail to reproduce 
the evolution of the $K$-band LF
and the $z=0$ gas-to-luminosity ratios 
(see L11 for a complete discussion of the impact of each 
SF law on the two models).

\subsection{The dependence of H$_2$/HI on galaxy morphology}

\begin{figure*}
\begin{center}
\includegraphics[width=1.0\textwidth]{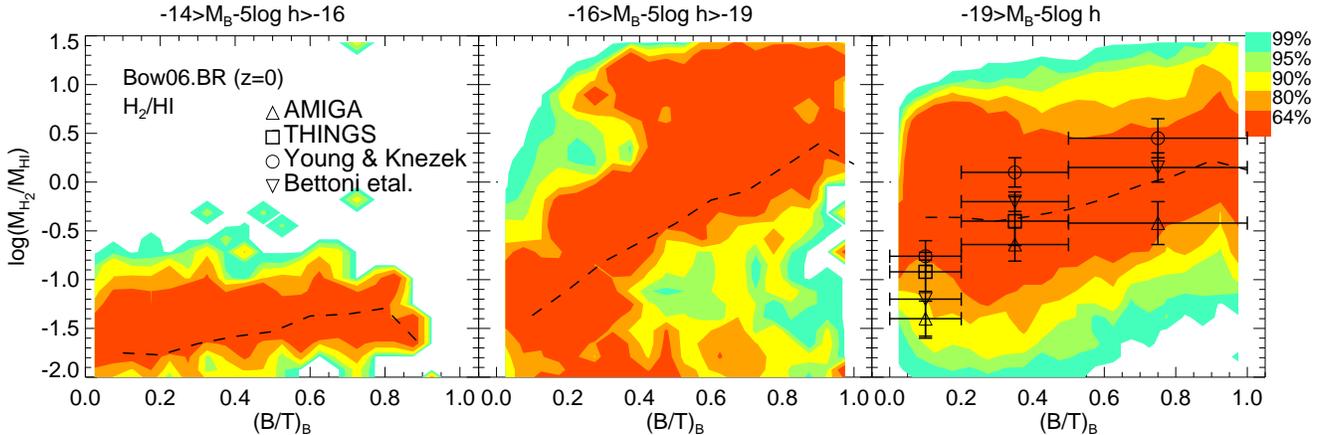}
\caption{{ Molecular-to-atomic hydrogen mass ratio, $M_{\rm H_2}/M_{\rm HI}$, as a function 
of the bulge-to-total luminosity in the $B$-band, $(B/T)_{B}$, in the Bow06.BR model, 
for galaxies with 
absolute $B$-band magnitudes, $M_B-5\rm \, log(h)$, in the range indicated on top of each panel. 
Contours
show the regions within which
different volume-weighted percentages of the galaxies lie for a
given $(B/T)_{B}$,
with the scale
shown by the key.
For reference, the dashed line shows the median of the model distribution. 
Observational results from \citet{Young89}, \citet{Bettoni03}, 
\citet{Leroy08} and \citet{Lisenfeld11} are shown as symbols 
in the right-hand panel, and we combine them so that
$B/T<0.2$ corresponds to Irr, Sm, Sd galaxies; $0.2<B/T<0.5$ corresponds 
to Sc, Sb, Sa galaxies; $B/T>0.5$ corresponds to E and S0 galaxies (see 
\citealt{deVaucouleurs91} for a description of each morphological type).}} 
\label{ScalingMstar2}
\end{center}
\end{figure*}

{ It has been shown observationally that the ratio of
H$_2$/HI masses correlates strongly with morphological type, with early-type 
galaxies characterised by higher H$_2$/HI ratios than late-type galaxies 
(e.g. \citealt{Young89}; \citealt{Bettoni03}; \citealt{Lisenfeld11}). 
Fig.~\ref{ScalingMstar2} shows the H$_2$/HI mass ratio as a function of the 
bulge-to-total luminosity ratio in the $B$-band, $B/T$, for different $B$-band 
absolute magnitude ranges for the Bow06.BR model. 
The right hand panel of Fig.~\ref{ScalingMstar2}, which shows the brightest 
galaxies, compares 
the model predictions with observations. All observational data have morphological types 
derived from a visual classification of the $B$-band images \citep{deVaucouleurs91}, and have 
also been selected in blue bands 
(e.g. \citealt{Simien86}; \citealt{Weinzirl09}). 
The comparison with observational data is shown for galaxies in the model with 
$M_B-5\rm \, log($$h\rm )<-19$, which roughly corresponds to the selection criteria 
applied in the observational data. 
Note that we have plotted only galaxies in the model that have 
$M_{\rm HI}/M_{\star}\ge 10^{-3}$ and $M_{\rm H_2}/M_{\star}\ge 10^{-3}$, which 
correspond to the lowest HI and H$_2$ gas fractions detected in 
the observational data shown. 
The Bow06.BR model predicts a relation between the H$_2$/HI ratio and $B/T$ 
that is in good agreement with the observations. 
Note that, for the last bin, $B/T<0.2$, the model predicts slightly higher 
median H$_2$/HI ratios than the values inferred from observations. However, in all cases
a constant CO$(1-0)$-H$_2$ conversion factor was assumed in the observational sample 
to infer the H$_2$ mass. 
This might not be a good approximation in the low-metallicity environments typical 
of irregular or late-type spirals, where the H$_2$ mass might be 
underestimated if a conversion factor typical of normal spiral galaxies is applied 
(e.g. \citealt{Boselli02}; see $\S 4.2.1$ for a discussion).}

{ The higher H$_2$/HI ratios in early-type galaxies can be understood in the context of 
the dependence of the H$_2$/HI ratio on gas pressure built into the BR SF law. 
Even though 
gas fractions in early-type galaxies are in general lower than in late-type galaxies, 
they are also systematically more compact than a late-type counterpart of the same mass 
(see \citealt{Lagos10}), resulting in higher gas pressure, and consequently a higher H$_2$/HI ratio. 
Also note that there is a dependence on galaxy luminosity: faint galaxies typically have  
lower H$_2$/HI ratios than their bright counterparts. This is due to the contribution 
of the stellar surface density to the pressure, which increases in massive galaxies 
(see Eq~\ref{Pext}).}

\subsection{The relation between HI, H$_2$ and stellar mass}

\begin{figure}
\begin{center}
\includegraphics[width=0.5\textwidth]{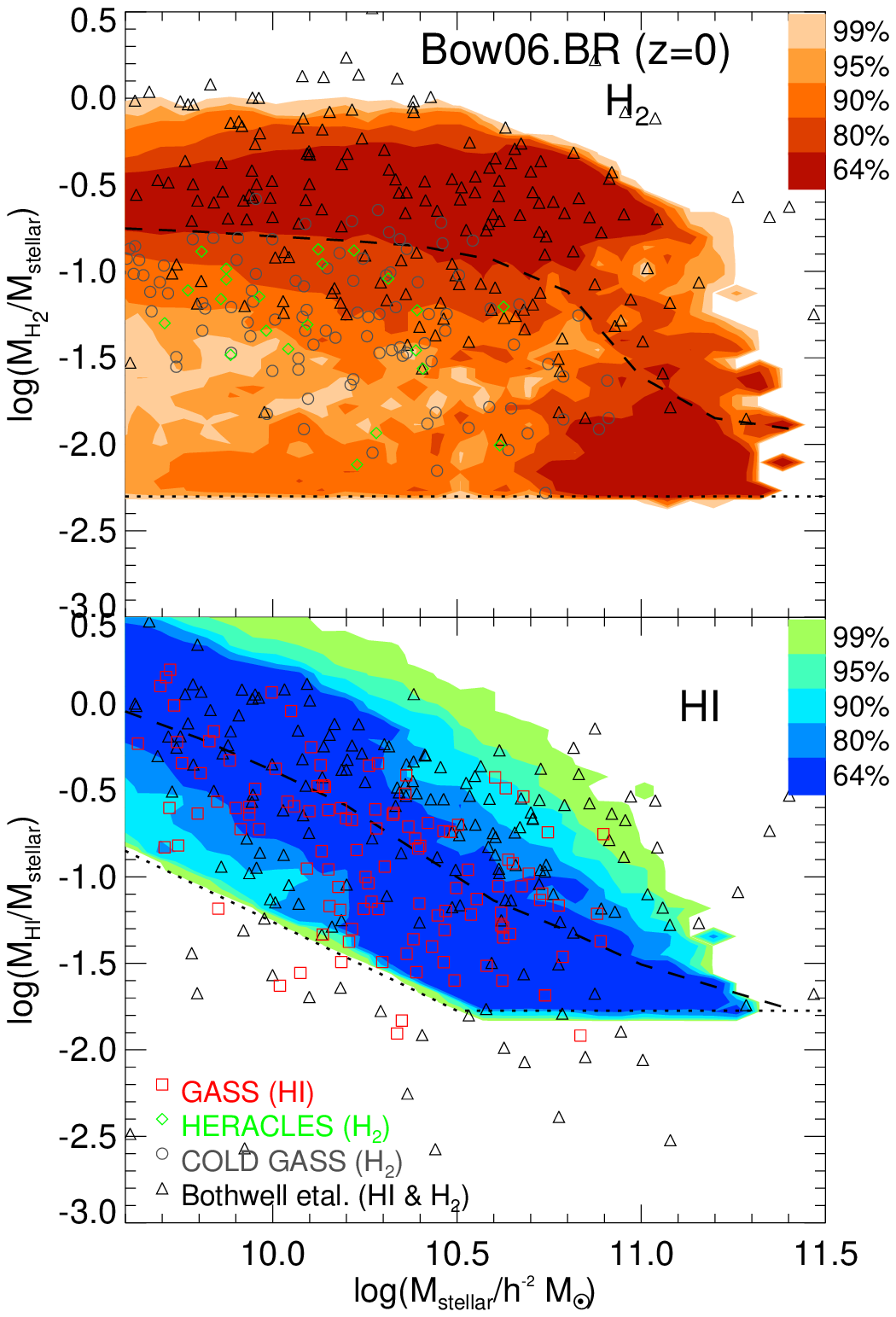}
\caption{{\it Top panel:} Molecular hydrogen-to-stellar mass ratio 
as a function of stellar mass for the Bow06.BR model at $z=0$. 
The horizontal dotted line shows an approximate sensitivity limit 
below which CO(1-0) is not detected in the different molecular surveys. 
Contours 
show the regions within which
different volume-weighted percentages of the galaxies lie for a 
given stellar mass and above the sensitivity limit, 
with the scale 
shown by the key. { For reference, the dashed line shows the median of the model 
distributions.} Observational data 
from the HERACLES survey \citep{Leroy09}, 
the COLD GASS 
survey \citep{Saintonge11} and the literature 
compilation of \citet{Bothwell09} are shown as symbols.
{\it Bottom panel:} As in the top panel 
but for the atomic hydrogen-to-stellar 
mass ratio. Observational 
data from the GASS catalogue \citep{Catinella10} and the literature compilation of 
Bothwell et al. are shown as symbols. The dotted line shows the HI 
sensitivity limit of the GASS catalogue.}  
\label{ScalingMstar}
\end{center}
\end{figure}

Another form of scaling relation often studied observationally 
is the atomic or molecular hydrogen-to-stellar mass ratio as 
a function of stellar mass. Recently, 
these relations have been reported for 
the atomic and molecular gas contents in 
a homogeneous sample of 
relatively massive galaxies 
by \citet{Catinella10} and \citet{Saintonge11}, respectively, with the aim 
of establishing fundamental relations 
between the stellar content of galaxies and their cold gas 
mass. Fig.~\ref{ScalingMstar} shows these relations for the Bow06.BR model at $z=0$ 
compared to values reported for individual galaxies 
in these surveys with detected H$_2$ or HI, respectively\footnote{Stellar masses
in the observational samples were inferred using a Chabrier IMF. We 
scale them by a factor $0.89$ to adapt them to our 
choice of a Kennicutt IMF.}. 
Also shown in the top panel  
are a data compilation from the 
literature presented in \citet{Bothwell09} and 
individual galaxies from the HERA CO Line Extragalactic Survey (HERACLES; \citealt{Leroy09}). 
Measurements of $M_{\rm H_2}/M_{\rm stellar}$ are subject 
to large errors (up to $0.25$~dex) given the 
uncertainty in the CO-H$_2$ conversion factor (see 
discussion in $\S 4.2.1$). Observed $M_{\rm HI}/M_{\rm stellar}$ ratios 
are more accurate given the direct detection of HI.     
Horizontal lines in Fig.~\ref{ScalingMstar} show representative
observational sensitivity limits (which are not exactly the same in all samples). 
Contours show the regions where different fractions of galaxies in the model 
lie, normalized in bins of stellar mass. The contours 
are not very sensitive to the location of the sensitivity limits.
The model predicts the right location and scatter of 
galaxies in these planes in contrast to previous 
models (see \citealt{Saintonge11} for a discussion).  
Note that the literature compilation of Bothwell et al. shows 
larger scatter than the HERACLES, GALEX Arecibo SDSS Survey (GASS) and 
CO Legacy Database for the GASS survey (COLD GASS). 
Saintonge et al. suggest that this is due to the inhomogeneity of 
 the literature compilation.

The $M_{\rm H_2}/M_{\rm stellar}$ ratio in the model is only weakly 
correlated with stellar mass, in contrast to 
the $M_{\rm HI}/M_{\rm stellar}$ ratio which is  
strongly dependent on stellar mass, in agreement  
with the observations. The 
decrease in the $M_{\rm HI}/M_{\rm stellar}$ ratio with 
increasing stellar mass is the dominant factor 
in determining the  
positive relation between 
H$_2$/HI and stellar mass in Fig.~\ref{Scaling1} 
given the small  
variations of the $M_{\rm H_2}/M_{\rm stellar}$ 
ratio with stellar mass. 

{ The predicted relation between $M_{\rm H_2}$ and $M_{\star}$ (top-panel of Fig.~\ref{ScalingMstar})
is close to linear for galaxies with $M_{\rm H_2}/M_{\star}\gtrsim 0.05$. These galaxies 
lie on the active star-forming sequence in the $M_{\star}-\rm SFR$ plane (see L11), 
due to an approximately linear relation between 
$M_{\rm H_2}$ and SFR, although characterised by a large scatter 
(see $\S 4.2.3$). What drives the approximately constant $\rm SFR/$$M_{\star}$ and 
$M_{\rm H_2}/M_{\star}$ ratios for galaxies on the active star-forming sequence is the balance 
between accretion and outflows, mainly regulated by the timescale for gas to be reincorporated 
into the host halo after ejection by SNe (see L11 for details). 
In the case of HI, the model predicts that, for
galaxies plotted in the bottom-panel of Fig.~\ref{ScalingMstar}, the 
HI weakly correlates with stellar mass, $M_{\rm HI}\propto M^{0.15}_{\star}$, 
as a result of the feedback mechanisms included in the model.}

\section{Atomic and molecular hydrogen mass functions}\label{MFs}

The two-phase ISM scheme implemented in {\texttt{GALFORM}} (see $\S 2$) 
allows us to study the evolution of the HI and $\rm H_2$ in 
galaxies in terms of the MFs and their evolution. In the next two subsections we 
analyse the main mechanisms which shape the HI and $\rm H_2$ MFs and investigate how 
these interplay to 
determine the model predictions.

\subsection{Atomic hydrogen mass function}

\begin{figure}
\begin{center}
\includegraphics[trim = 7mm 5.5mm 2mm 2mm,clip,width=0.48\textwidth]{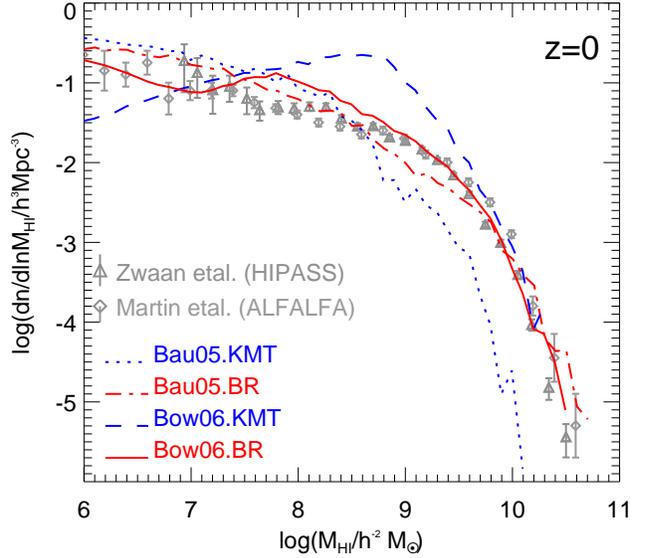}
\caption{The HI mass function at $z=0$ for the Bau05.KMT (dotted line), Bau05.BR (dot-dashed line), 
Bow06.KMT (dashed line) and Bow06.BR (solid line) models. 
Symbols show the $z=0$ observational estimates from \citet{Zwaan05} 
using HIPASS and from \citet{Martin10} using ALFALFA, as 
labelled.}
\label{HIz0}
\end{center}
\end{figure}

Fig.~\ref{HIz0} shows the $z=0$ HI mass functions (HI MF) 
for the Bow06 and Bau05 models using the BR and the 
KMT SF laws, and observational 
results from \citet{Zwaan05} and \citet{Martin10}. 

With the BR SF law, both
the Bow06 and Bau05 models  
give predictions which are in reasonable agreement with the observed HI MF.
However, when 
applying the KMT SF law both models give  
a poor match to the observed HI MF, with the Bau05 model 
underpredicting the number density 
of massive HI galaxies, whilst the Bow06 model greatly  
overpredicts the abundance of galaxies around 
$M_{\rm HI}\sim 10^9 h^{-2} \rm M_{\odot}$. 
The latter is expected from the poor agreement between the 
H$_2$/HI-stellar mass scaling relation predicted by Bow06.KMT 
and the observed one (Fig.~\ref{Scaling1}). In the case of Bau05.KMT, the poor 
agreement is expected from the low   
gas-to-luminosity ratios reported by L11. 

{ \citet{Cook10} showed that, by including 
the BR SF law self-consistently in their semi-analytic model, the number density of low 
HI mass galaxies is substantially increased, in agreement with our results. However, in their 
model this effect is still not enough to bring the predictions into agreement with 
the observed HI MF. The overproduction of low mass galaxies is also seen in their predicted optical LFs, 
suggesting that the difference resides in the treatment of SN feedback and reionisation 
(see \citealt{Bower06}; \citealt{Benson10}). 
In the case of the KMT SF law, there are no published results using the full SF law in semi-analytical models. 
\citet{Fu10} implemented the H$_2$/HI metallicity- and gas surface density-dependence of the KMT SF law, 
but assumed a constant 
SF timescale for the molecular gas, instead of the original 
dependence on gas surface density (see Eq.~\ref{sfreq2}). Furthermore, these authors compared their predictions 
to the observations of the HI MF only over the restricted range 
$10^9 \lesssim M_{\rm HI}/h^{-2} M_{\odot} \lesssim 10^{10}$, one decade in mass compared to the 
five decades plotted in Fig.~\ref{HIz0}, so that 
it is hard to judge how well this KMT-like model really performs.}

Given that neither of the models using  
the KMT SF law predict the 
right HI MF or H$_2$/HI scaling relations, and that the Bau05 model fails to predict 
the right $K$-band LF at $z>0.5$, we 
now focus on the Bow06.BR model when presenting the 
predictions for the gas contents of galaxies at low and high redshifts. 

\subsubsection{The composition of the HI mass function} 

\begin{figure}
\begin{center}
\includegraphics[trim = 4mm 2mm 2mm 2mm,clip,width=0.49\textwidth]{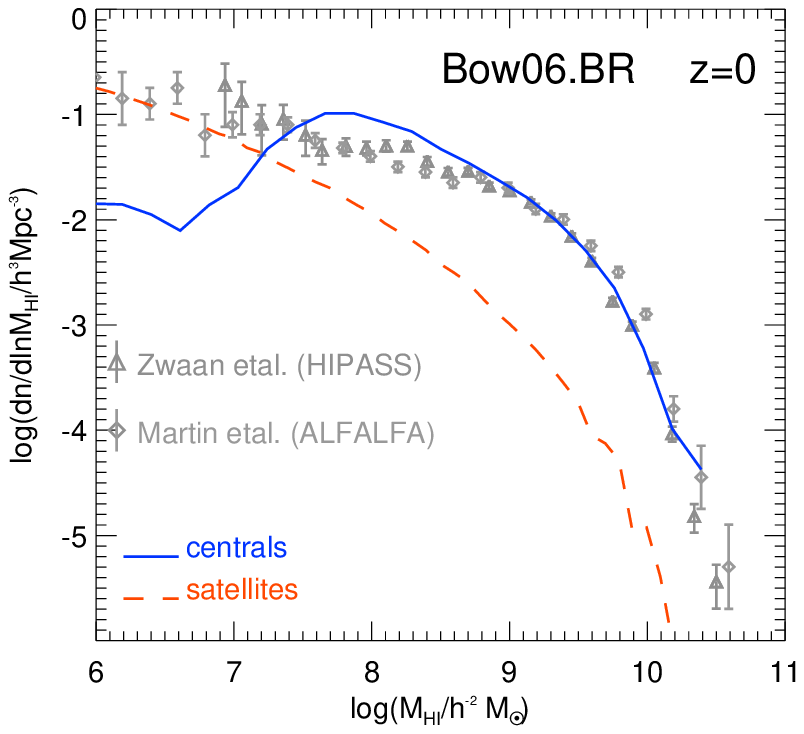}
\includegraphics[trim = 4mm 2mm 2mm 2mm,clip,width=0.49\textwidth]{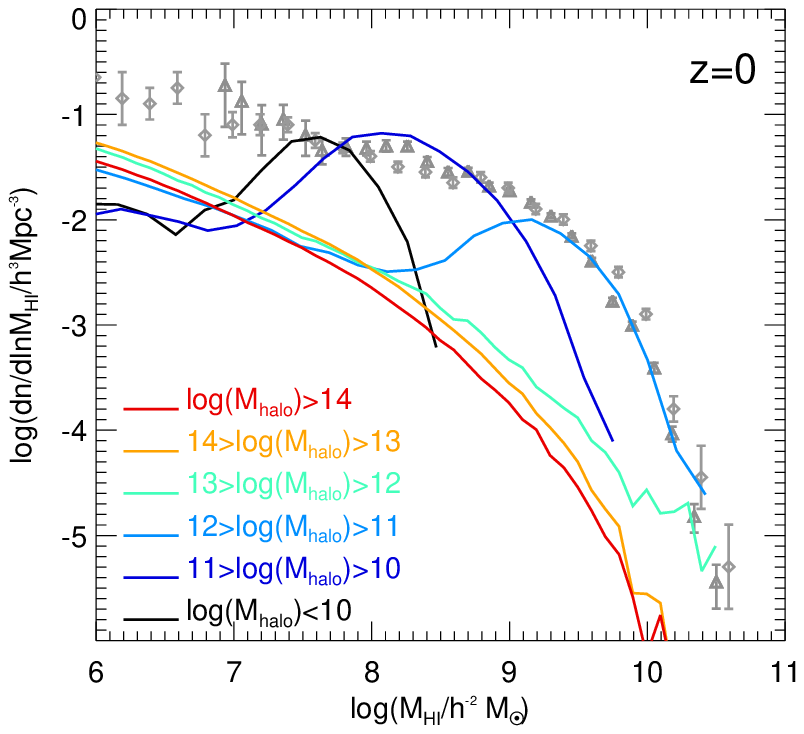}
\caption{The HI mass function at $z=0$ in the Bow06.BR model, 
distinguishing between the contribution from central (solid line) and satellite (dashed line) 
galaxies in the top panel, and from galaxies hosted by DM halos of 
different masses, as labelled, in the bottom panel. }
\label{HIz0sats}
\end{center}
\end{figure}

The low mass end of the $z=0$ HI MF ($M_{\rm HI}\lesssim 10^7 h^{-2} \rm M_{\odot}$), 
is dominated by satellite galaxies, as can be seen in the top panel of
Fig.~\ref{HIz0sats}. Central galaxies contribute less to the number density 
in this mass range due to reionisation. Hot gas in halos with circular velocity  
$V_{\rm circ}<30 \,\rm km\,s^{-1}$ is not allowed to 
cool at $z<10$, thereby suppressing the accretion of 
cold gas onto central galaxies hosted by these halos. 
These haloes have masses typically $M_{\rm halo}<10^{11} \,h^{-1} M_{\odot}$ 
\citep{Benson10b}. 
This suggests that the HI content of galaxies in 
isolation may lead to constraints on reionisation (Kim et al. in prep.). 
The HI MF of satellite galaxies does not show the dip at low HI masses observed in  
the HI MF of central galaxies because the satellites 
were mainly formed before reionisation.
Galaxies of intermediate 
and high HI mass mainly correspond to central galaxies. 
The predominance in the HI MF of central and satellite 
galaxies in the high and low mass 
ends, respectively, is independent of the model adopted. 

The Bow06.BR model slightly overpredicts the number density of galaxies in the mass range 
$3\times 10^7 h^{-2} \rm M_{\odot}<M_{\rm HI}<3\times 10^8$$h^{-2} \rm M_{\odot}$. 
This is mainly due to the slightly larger radii of the model galaxies 
compared to observations (see L11), which leads
 to lower pressure within the galactic disk and therefore 
to a slight overestimate of the atomic hydrogen content.

The bottom panel of Fig.~\ref{HIz0sats} shows the contribution 
to the HI MF from galaxies hosted by halos of different masses.
The HI MF at intermediate and high HI 
masses, i.e. $M_{\rm HI}\gtrsim 5\times 10^7$$h^{-2} \rm M_{\odot}$, 
is dominated by galaxies hosted by low and intermediate
mass DM halos,
$M_{\rm halo}\lesssim 10^{12} h^{-1} \rm M_{\odot}$, while lower HI mass 
galaxies are primarily satellites in higher mass halos.  
This scale in the DM halo 
mass ($M_{\rm halo} \approx 10^{12} h^{-1} \rm M_{\odot}$) has been
shown to be set by the efficient suppression of SF in higher mass DM haloes,  
mainly driven by AGN feedback which shuts down gas cooling \citep{Kim10}. 
In lower mass halos, in which AGN feedback does not 
suppress gas cooling, the cold gas 
content scales with the stellar mass of the galaxy 
and with the mass of the host DM halo.

\subsubsection{Evolution of the HI mass function}

\begin{figure}
\begin{center}
\includegraphics[trim = 5mm 3.5mm 2mm 2mm,clip,width=0.49\textwidth]{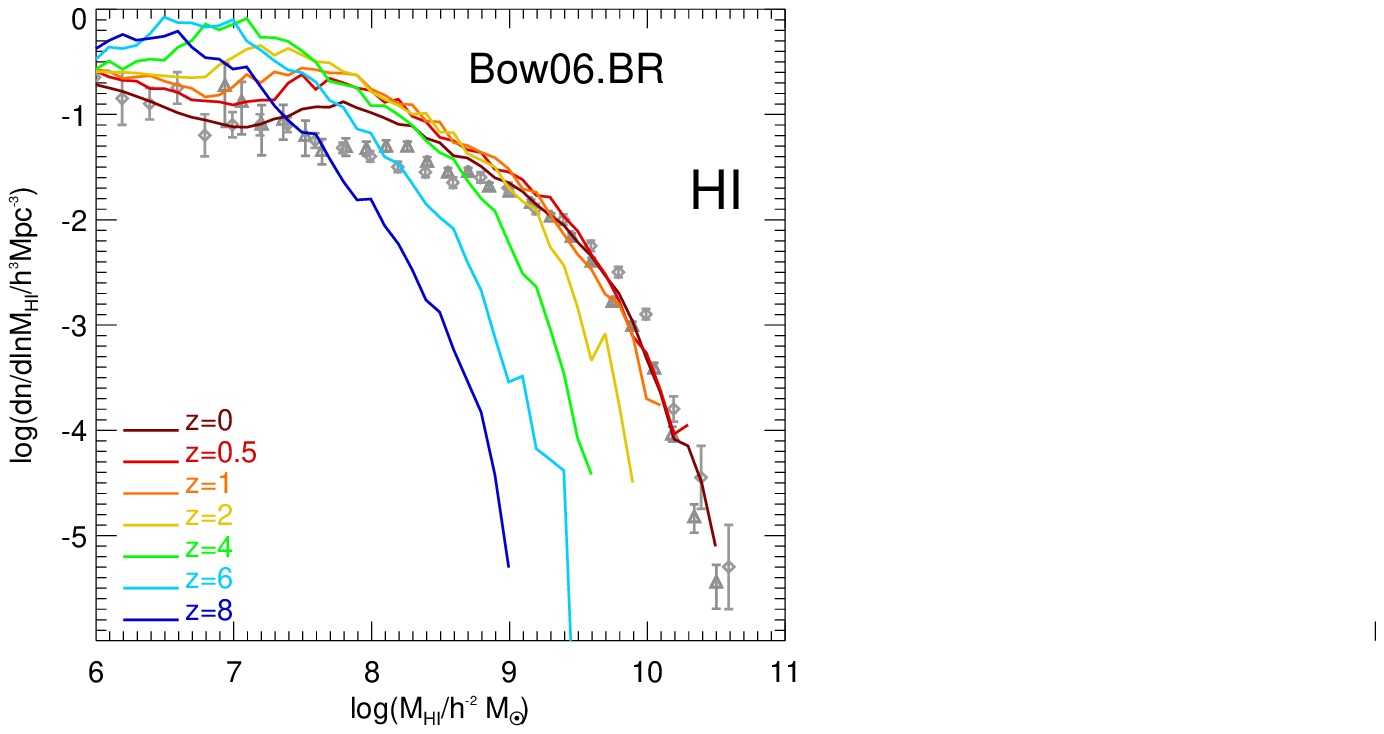}
\caption{
The HI mass functions for the Bow06.BR model at different redshifts, as labelled.
For reference, 
we show the $z=0$ observational results from \citet{Zwaan05} and \citet{Martin10}.}
\label{HIevo}
\end{center}
\end{figure}

Fig.~\ref{HIevo} shows the evolution 
of the HI MF from $z=8$ to $z=0$.
There is a high abundance of low HI mass galaxies 
at high redshift. This 
reduces with declining redshift as gas is depleted mainly through 
quiescent SF and starbursts.
The number density of galaxies
with low HI masses, $M_{\rm HI}\lesssim 10^{7} h^{-2} \rm M_{\odot}$, 
increases with redshift, being a
factor $4$ larger at $z=6$ than at $z=0$.
The number density of HI galaxies with masses  
$10^{7} h^{-2} \rm M_{\odot} \lesssim M_{\rm HI}\lesssim 10^{8} {\it h}^{-2} \rm M_{\odot}$ 
increases by a factor of $2$ from $z=0$ to $z=1$, with little evolution 
to $z=4$. At $z>4$, the number density of galaxies in this mass range drops again.  
The high-mass end of the HI MF, $M_{\rm HI}\gtrsim 5\times 10^{8} h^{-2} \rm M_{\odot}$, 
grows hierarchically 
until $z\approx1$, increasing in number 
density by more than two orders of magnitude. 
The evolution of the high-mass end tracks the formation 
of more massive haloes 
in which gas can cool, until AGN heating 
becomes important (at $z\approx1$).
At $z<1$ 
the high-mass end does not show appreciable evolution.
{ The HI MF from the break upwards in mass is dominated          
by galaxies in intermediate mass halos. These are less affected by AGN feedback,  
driving the hierarchical growth in the number density in the high HI mass range.
Higher mass halos, $M_{\rm halo}\gtrsim 10^{12}\, h^{-1} M_{\odot}$, are subject to AGN feedback, which suppresses 
the cooling flow and reduces the cold gas reservoir in the 
central galaxies of these haloes. Lower mass haloes, $M_{\rm halo}\lesssim 10^{11}\, h^{-1} M_{\odot}$, 
are susceptible 
to SNe feedback, which depletes the cold gas supply by heating the gas and 
returning it to the hot halo.}

Note that our choice of parameters 
for reionisation affects mainly the low mass end of the MF 
at high redshifts. At $z=0$ the abundance of galaxies with 
$M_{\rm HI}\lesssim 10^{8} h^{-2} \rm M_{\odot}$ would be lower 
by a factor of $2$ if a lower  
photoionisation redshift cut of $z_{\rm reion}=6$ was assumed. 

\begin{figure}
\begin{center}
\includegraphics[trim = 1mm 3mm 0mm 2mm,clip,width=0.49\textwidth]{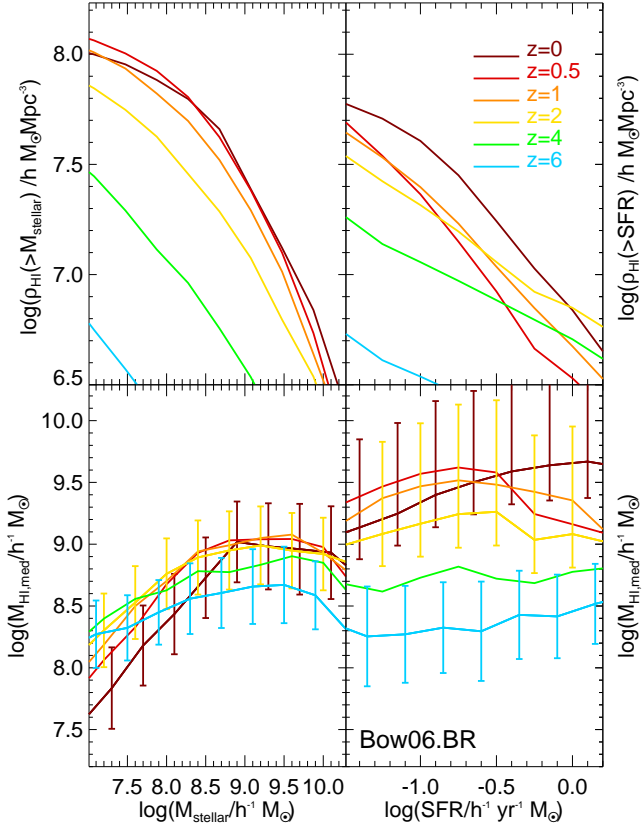}
\caption{{\it Top panels:} cumulative HI mass density 
for galaxies in the Bow06.BR model with stellar 
masses (left panel) or SFRs (right panel) larger than a given value 
at different redshifts, as labelled. {\it Bottom panels:} median HI mass of galaxies 
as a function of stellar mass (left panel) and SFR (right panel).
Errorbars correspond to the $10$ and $90$ percentiles of the distributions,
and are shown for three different redshifts.}
\label{HIevo2}
\end{center}
\end{figure}

The characterisation of the HI MF at redshifts higher than $z\approx 0$ will be a 
difficult task in future observations. The first measurements will come 
from stacking of stellar mass- or SFR-selected galaxy samples, 
as has been done in the local Universe (e.g. \citealt{Verheijen07}; \citealt{Lah07}, 2009).
Fig.~\ref{HIevo2} shows the 
cumulative HI mass per unit volume, $\rho_{\rm HI}$, at different redshifts 
for samples of galaxies in the Bow06.BR model that have stellar masses 
or SFRs larger than $M_{\rm stellar}$ or $\rm SFR$, respectively. 
The estimated $\rho_{\rm HI}$ from source stacking 
can be directly compared to our predictions 
for the same lower stellar mass or SFR limit. 
The median HI mass of galaxies as a function of 
stellar mass and SFR, 
is shown in the 
bottom-panels of Fig.~\ref{HIevo2}. 
Errorbars represent the $10$ and $90$ percentiles of the distributions, 
and are shown for three different redshifts.
The HI mass-stellar mass relation 
becomes shallower with increasing redshift, but the 
scatter around the median depends only weakly on redshift. 
The turnover of the median HI mass at 
$M_{\rm stellar}\gtrsim 10^{9} h^{-2}\, M_{\odot}$ 
is produced by AGN feedback that efficiently suppresses any 
further gas cooling onto massive galaxies and, consequently, their 
cold gas content is reduced. 
The HI mass is weakly correlated with SFR, particularly at 
high redshift. The median HI mass of galaxies spanning SFRs in the range 
plotted decreases with increasing redshift. This suggests 
that in order to detect low HI masses in observations 
at $z\approx 0.5-2$, stellar mass selected samples should be 
more effective than SFR selected samples. However, it would still be necessary 
to sample down to very low stellar masses 
($M_{\rm stellar}>10^{8}\, h^{-1}\rm M_{\odot}$).
Upcoming HI surveys using telescopes such as ASKAP and MeerKAT, 
will be able to probe down to these HI masses.

\subsection{Molecular hydrogen mass function}

The cold gas content is affected by SF, feedback processes, 
accretion of new cooled gas, but also by the evolution of galaxy sizes, 
given that our prescriptions to calculate the H$_2$ abundance depend 
explicitly on the gas density (see $\ 2.3$).
Our aim is to disentangle which processes 
primarily determine the evolution of H$_2$ in galaxies.

\subsubsection{The present-day $\rm CO(1-0)$ luminosity function}

Observationally, the most commonly 
used tracer of the H$_2$ molecule is the CO molecule, and in particular, 
the $\rm CO(1-0)$ transition which is emitted in the densest, 
coldest regions of the ISM, where 
the H$_2$ is locked up. 
Given that in the model we estimate the H$_2$ content, 
we use a conversion factor to estimate the 
$\rm CO(1-0)$ emission for a given abundance of H$_2$,

\begin{figure}
\begin{center}
\includegraphics[trim = 5mm 3.5mm 2mm 2mm,clip,width=0.5\textwidth]{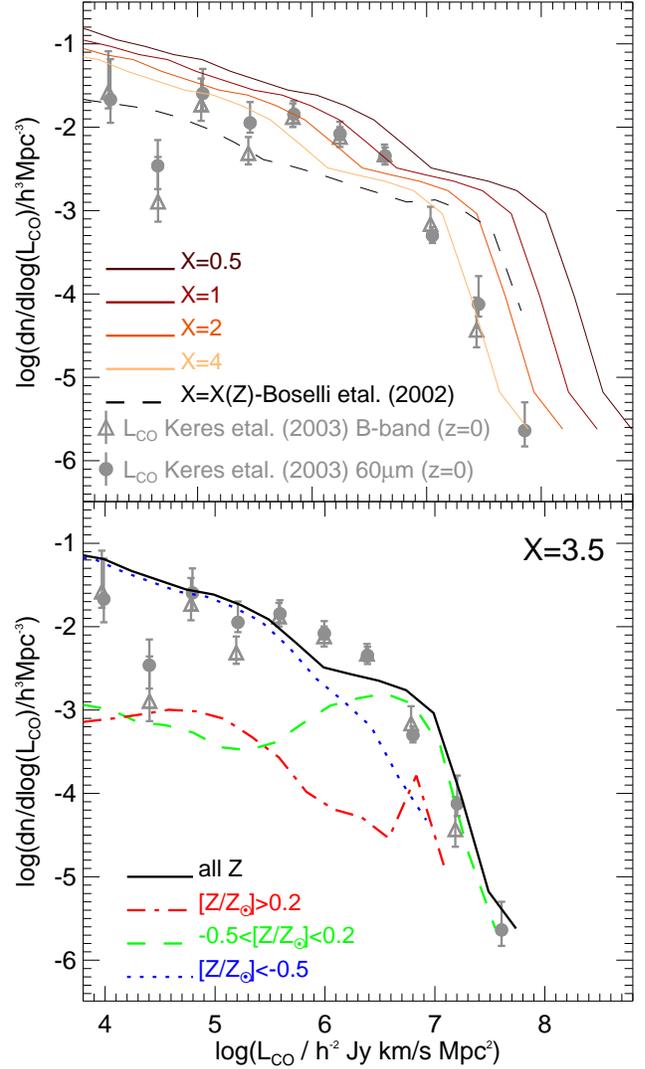}
\caption{The $\rm CO(1-0)$ luminosity function at $z=0$ 
for the Bow06.BR model compared to the observational estimates of \citet{Keres03} 
for $B$-band (triangles) and a $60\,\mu$m (filled circles) selected samples of galaxies. 
To calculate the $\rm CO(1-0)$ luminosities
we assume a $\rm H_2$-to-CO conversion factor, $\rm X$, where  
$N_{\rm H_2}/\rm cm^{-2}=X\times10^{-20}\, I_{\rm CO}/\rm K\,km\,s^{-1}$. 
In the top panel, solid lines show the model predictions 
using fixed values of X, as labelled, and the dashed line shows the  
LF using the metallicity-dependent 
conversion factor, $\rm X($$Z)$, of \citet{Boselli02}. In the bottom panel, 
the solid line shows the LF 
resulting from a conversion factor $X=3.5$ 
and the different lines show the contribution 
of galaxies in different metallicity ranges, as labelled, for this choice of X.}
\label{LFCOz0}
\end{center}
\end{figure}

\begin{equation}
 I_{\rm CO}/\rm K\,km\,s^{-1}=\frac{{\it N}_{\rm H_2}/\rm cm^{-2}}{\rm X\times10^{-20}}.
\label{Xfac}
\end{equation}    

\noindent Here $N_{\rm H_2}$ is the column density of H$_2$ and 
$I_{\rm CO}$ is the integrated CO$(1-0)$ line intensity per unit 
surface area. The value of $X$ has been inferred observationally in a few 
galaxies, mainly through virial estimations.
Typical estimates for normal spiral galaxies range 
between $X\approx 2.0-3.5$ (e.g. 
\citealt{Young91}; \citealt{Boselli02}; \citealt{Blitz07}). 
However, systematic variations in the value of $X$ are both,  
theoretically predicted and inferred observationally.
For instance, theoretical calculations predict that 
low metallicities, characteristic of dwarf galaxies, or   
high densities and large optical depths of molecular clumps in starburst galaxies, 
should change X by a factor of up to $5-10$ in either direction 
(e.g. \citealt{Bell07}; \citealt{Meijerink07}; \citealt{Bayet09}). 
Observations of dwarf galaxies favour 
larger conversion factors 
(e.g. $X\approx 7$; e.g. \citealt{Boselli02}), while the opposite holds  
in starburst galaxies (e.g. $X\approx 0.5$; e.g. \citealt{Meier04}). 
This suggests a metallicity-dependent conversion factor X. 
However, 
estimates of the 
correlation between X and 
metallicity 
in nearby galaxies  
vary significantly, 
from finding no correlation, 
when virial equilibrium of giant molecular clouds (GMCs) 
is assumed (e.g. \citealt{Young91}; \citealt{Blitz07}),  
to correlations as strong as $X\propto (Z/Z_{\odot})^{-1}$, when 
the total gas content is inferred from the dust content on assuming 
metallicity dependent dust-to-gas ratios   
(e.g. \citealt{Guelin93}; \citealt{Boselli02}).

We estimate the $\rm CO(1-0)$ LF by using different conversion 
factors favoured by the observational estimates described above.  
The top panel of Fig.~\ref{LFCOz0} shows the $\rm CO(1-0)$ LF at $z=0$ when 
different constant conversion factors are assumed (i.e. independent of 
galaxy properties; solid lines).
Observational estimates of the $\rm CO(1-0)$ LF, made using 
a $B$-band and a $60\,\mu$m selected sample, are plotted 
using symbols \citep{Keres03}.
The model
slightly underestimates 
the number density at $L^{\star}$ for $\rm X>1$,  
but gives good agreement at fainter and at 
brighter luminosities for sufficiently large values of X (such as the ones 
inferred in normal spiral galaxies). 
In the predicted $\rm CO(1-0)$ LF we include all galaxies 
with a $L_{\rm CO}>10^3 \, \rm Jy \,km/s\,Mpc^{2}$, while the 
LFs from Keres et al. were  
inferred from samples of 
galaxies selected using $60\,\mu \rm m$ or $B$-band 
fluxes. These criteria might bias the LF 
towards galaxies with large amounts of dust or 
large recent SF. 
More data is needed from blind CO surveys 
in order to characterise the CO LF 
in non-biased samples of galaxies. This will 
be possible with new instruments such as 
the LMT.

In order to illustrate how much our predictions for the 
$\rm CO(1-0)$ LF at $z=0$ vary when 
adopting a metallicity-dependent conversion factor, $X(Z)$, 
inferred independently from observations, 
we also plot in the top panel of Fig.~\ref{LFCOz0}  
the LF when the $X(Z)$ relation from \citet{Boselli02} is adopted (dashed line),
$\rm log({\it X})=0.5_{-0.2}^{+0.2}-1.02_{-0.05}^{+0.05}\rm log(\rm Z/Z_{\odot})$.
Note that this correlation was determined 
using a sample of $12$ galaxies with $\rm CO(1-0)$ 
luminosities in quite a narrow range, 
$\rm L_{\rm CO}\approx 5\times 10^5-5\times 10^6$~$\rm Jy\, km/s\, Mpc^{2}$. 
On adopting this conversion factor, 
the model largely underestimates the break of the LF. 
This is due to the contribution of galaxies with different metallicities 
to the $\rm CO(1-0)$ LF, as shown 
in the bottom panel of Fig.~\ref{LFCOz0}. 
The faint-end is dominated by low-metallicity galaxies 
($\rm Z<Z_{\odot}/3$), while 
high-metallicity galaxies ($\rm Z>Z_{\odot}/3$) dominate the bright-end. 
A smaller $X$ for low-metallicity 
galaxies combined with a larger $X$ for high-metallicity galaxies 
would give better agreement with the observed data. However, such 
a dependence of $X$ on metallicity is opposite 
to that inferred by Boselli et al.    

By considering a 
dependence of $X$ on metallicity alone  
we are ignoring possible variations with other physical properties 
which could influence the state of 
GMCs, such as the interstellar 
far-UV radiation field and variations in the 
column density of gas (see for instance \citealt{Pelupessy06}; 
\citealt{Bayet09}; \citealt{Pelupessy09}; 
\citealt{Papadopoulos10}). { Recently, \citet{Obreschkow09d} showed that 
by using a simple phenomenological model to calculate the luminosity of different 
CO transitions, which includes information about the 
ISM of galaxies, the CO$(1-0)$ luminosity function of Fig.~\ref{LFCOz0} can be 
reproduced. However, this modelling introduces several 
extra free parameters into the model which, in most cases, are not well constrained 
by observations.}
A more detailed 
calculation of the $\rm CO$ LF which 
takes into account the characteristics of the local ISM environment 
is beyond the scope of this paper and is addressed in 
a forthcoming paper (Lagos et al. 2011, in prep.). 

For simplicity, in the next subsection we use a fixed $\rm CO(1-0)$-H$_2$ 
conversion factor of $X=3.5$ 
for galaxies undergoing quiescent SF and $X=0.5$ 
for those experiencing starbursts. In the case of galaxies 
undergoing both SF modes, we use $X=3.5$ and $X=0.5$ for the quiescent 
and the burst components, respectively, following the discussion above. 

\subsubsection{Evolution of the H$_2$ mass function}
\begin{figure}
\begin{center}
\includegraphics[trim = 5mm 3.2mm 2mm 2mm,clip,width=0.5\textwidth]{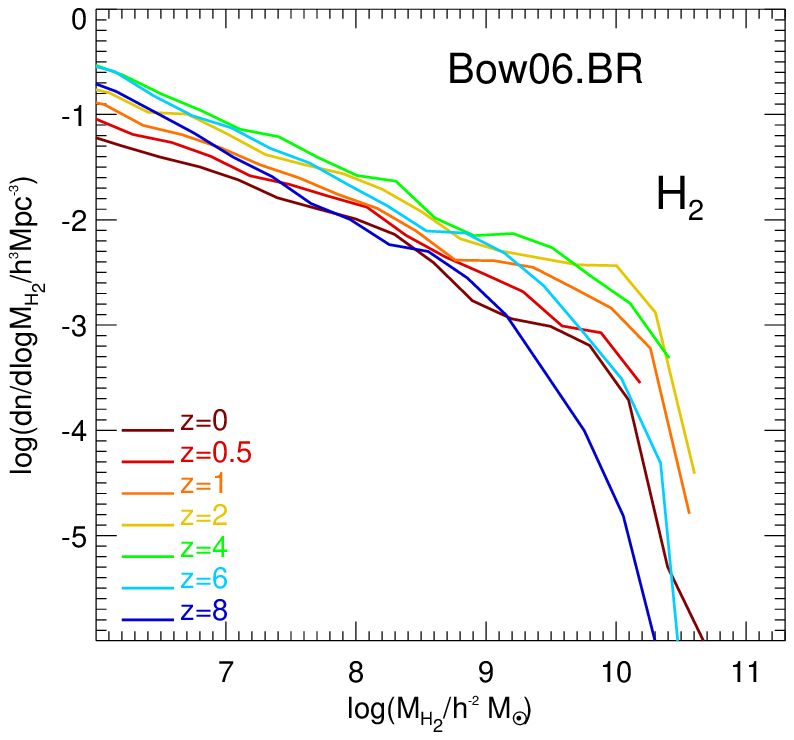}
\includegraphics[trim = 5mm 3.2mm 2mm 2mm,clip,width=0.5\textwidth]{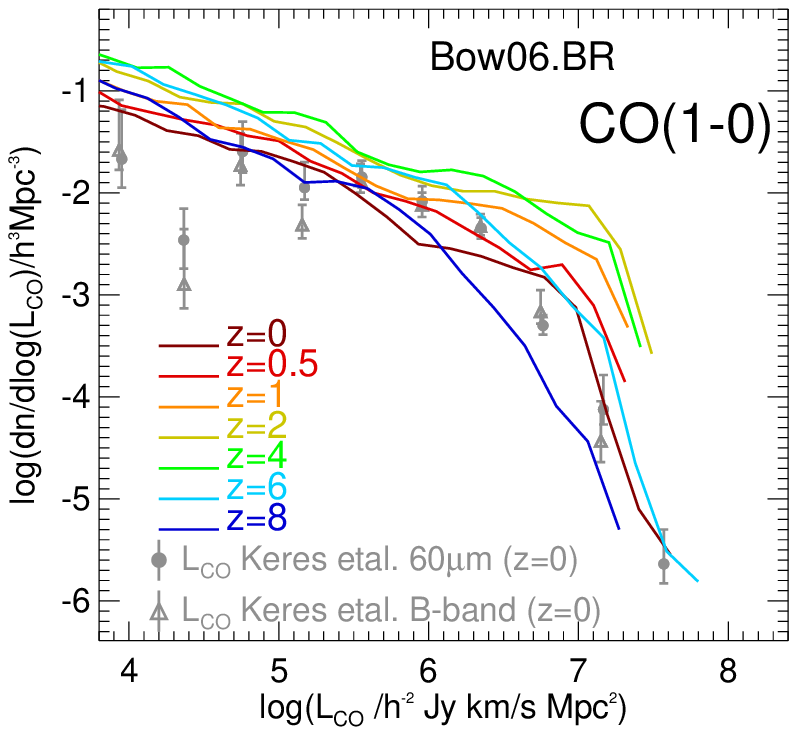}
\caption{The H$_2$ MF (top panel) and the $\rm CO(1-0)$ LF (bottom panel) at different 
redshifts, as labelled, for the Bow06.BR model. Two fixed $\rm CO(1-0)$-H$_2$ conversion factors 
are assumed: $X=3.5$ for quiescent SF and $X=0.5$ for burst SF. 
For reference, we show the $\rm CO(1-0)$ LFs estimated  
at $z=0$ by \citet{Keres03} from $B$-band 
and $60\,\mu$m-selected galaxy samples.}
\label{H2evo}
\end{center}
\end{figure}

At high-redshift, measurements of the $\rm CO(J\rightarrow J-1)$ LF are not 
currently available. 
ALMA will, however, provide measurements 
of molecular emission lines in high 
redshift galaxies with high accuracy. 
We therefore present 
in Fig.~\ref{H2evo} predictions for the H$_2$ MF and $\rm CO(1-0)$ LF up to $z=8$. 

The high-mass end of the 
H$_2$ MF shows 
strong evolution from $z=8$ to $z=4$, with the number density of galaxies increasing 
by an order of magnitude. In contrast, the number density of low H$_2$ mass galaxies 
stays approximately constant over the same redshift range. 
From $z=4$ to $z=2$ the H$_2$ MF hardly evolves, 
with the number density of galaxies remaining the same over the 
whole mass range. From $z=2$ to $z=0$, the number density of massive galaxies 
decreases by an order of magnitude, while the low-mass end decreases only 
by a factor $\approx 3$. 
The peak in the number density of massive H$_2$ galaxies at $z=2-3$ 
coincides with the peak of the SF activity 
(see L11; \citealt{Fanidakis10b}), 
in which huge amounts of H$_2$ are consumed forming new stars. 
The following decrease in the number density at $z<2$ overlaps 
with strong galactic size evolution, where galaxies at lower 
redshift are systematically larger than their high redshift counterparts, 
reducing the gas surface density and therefore, the H$_2$ fraction. 
We return to this point in $\S$\ref{ScalingRels}. 
The peak in the number density 
of high H$_2$ mass galaxies at $z=2-3$ and the following decrease, 
contrasts with the monotonic increase 
in the number density of high HI mass galaxies with time, suggesting 
a strong evolution of the H$_2$/HI global ratio with redshift. 
We come back to this point in $\S 6$. 

The evolution of the $\rm CO(1-0)$ LF with redshift is shown 
in the bottom panel of Fig.~\ref{H2evo}. Note that the $z=0$ 
LF is very similar to the one shown in the bottom 
panel of Fig.~\ref{LFCOz0}, in which we assume a 
fixed CO-H$_2$ conversion $X=3.5$ 
for all galaxies. This is due to the low number of starburst 
events at $z=0$. However, at higher redshifts, starbursts 
contribute more to the LF; the most luminous events at high redshift, in terms of CO 
luminosity, correspond to 
starbursts (see $\S 4.2.3$). 

Recently 
\citet{Geach11} compared the evolution of the observed 
molecular-to-(stellar plus molecular mass) ratio (which 
approximates to the gas-to-baryonic ratio if the HI fraction is small), 
$f_{\rm gas}=M_{\rm mol}/(M_{\rm
stellar}+M_{\rm mol})$, with predictions for the Bow06.BR model { at $z\le2$},
and found that the model 
gives a good match to the observed $f_{\rm gas}$ evolution  
after applying the same observational selection cuts.

\subsubsection{The IR-CO luminosity relation}\label{FIRCO}

One way to study the global relation between the SFR and the molecular
gas mass in galaxies, and hence to constrain the SF law, is through
the relation between the total IR luminosity, $L_{\rm IR}$, and the
CO$(1-0)$ luminosity, $L_{\rm CO}$. We here define the total IR
luminosity to be the integral over the rest-frame wavelength range
$8-1000 \mu$m, which approximates the total luminosity emitted by
interstellar dust, free from contamination by starlight. In dusty
star-forming galaxies, most of the UV radiation from young stars is
absorbed by dust, and this dominates the heating of the dust, with
only a small contribution to the heating from older stars. Under these
conditions, and if there is no significant heating of the dust by an
AGN, we expect that $L_{\rm IR}$ should be approximately proportional
to the SFR, with a proportionality factor that depends mainly on the
IMF. On the other hand, as already discussed, the CO$(1-0)$ luminosity
has been found observationally to trace the molecular gas mass in
local galaxies, albeit with a proportionality factor that is different
in starbursts from quiescently star-forming galaxies. Observations
suggest that sub-millimeter galaxies (SMGs) and QSOs at high redshift
lie on a similar IR-CO luminosity relation to luminous IR galaxies
(LIRGs) and ultra-luminous IR galaxies (ULIRGs) observed in the local
Universe (see \citealt{Solomon05} for a review). We investigate here
whether our model predictions are consistent with these observational
results.

\begin{figure}
\begin{center}
\includegraphics[trim = 2mm 1mm 1mm 1mm,clip,width=0.5\textwidth]{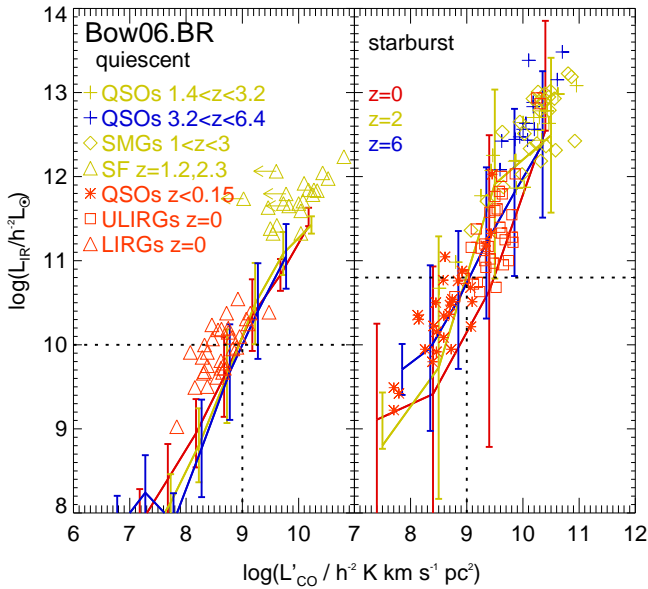}
\caption{Infrared luminosity as a function of the CO($1-0$) luminosity
for the Bow06.BR model at $z=0$ (red lines), $z=2$ (yellow lines) and
$z=6$ (blue lines) for quiescent (left panel) and starburst galaxies
(right panel). Solid lines show the median, while errorbars show the
$10$ and $90$ percentiles of the distributions.  To aid the comparison
between the quiescent and starburst galaxies, dotted lines show the
predicted median IR luminosity at $z=2$ for a CO luminosity of
$10^9$~$h^{-2}\, \rm K\, km\, s^{-1} \,pc^{2}$.  We also show the following
observational data: local LIRGs from \citet{Gao04} (triangles); local
ULIRGs from \citet{Solomon97} (squares); QSOs at $z\lesssim 0.15$ from
\citet{Scoville03}, \citet{Evans06} and \citet{Bertram07} (asterisks);
star-forming galaxies at $z\approx 1.2$ and $z\approx 2.3$ from
\citet{Tacconi10} and \citet{Genzel10} (triangles); SMGs at $1\lesssim
z\lesssim 3$ from \citet{Greve05}, \citet{Solomon05} and
\citet{Tacconi06} (diamonds); and QSOs at $1.4\lesssim z\lesssim 6.4$
from \citet{Riechers10} (crosses).  Note that most of the
observational data on the CO$(1-0)$ luminosity at high redshifts are
inferred from the luminosities of higher CO transitions rather than
being directly measured.  }
\label{FIRCOlums}
\end{center}
\end{figure}

We show in Fig.~\ref{FIRCOlums} the predicted IR-CO$(1-0)$ luminosity
relation, $L_{\rm IR}-L'_{\rm CO}$, for the Bow06.BR model at
different redshifts, compared to observational data for different
types of galaxies. The predicted CO$(1-0)$ luminosities for the model
galaxies are calculated from their $H_2$ masses as in \S4.2.1, using
conversion factors $X=3.5$ for quiescent galaxies and $X=0.5$ for
bursts. To facilitate the comparison with observations, we here
express the CO luminosities $L'_{\rm CO}$ in units of $\rm K\, km\,
s^{-1} \,pc^{2}$, which corresponds to expressing the CO line
intensity as a brightness temperature.  We predict the IR luminosities
of the model galaxies using the method described in \citet{Lacey11}
and \citet{Gonzalez10} (see also \citet{Lacey11b}), which uses a
physical model for the dust extinction at each wavelength to calculate
the total amount of stellar radiation absorbed by dust in each galaxy,
which is then equal to its total IR luminosity. The dust model assumes
a two-phase interstellar medium, with star-forming clouds embedded in
a diffuse medium. The total mass of dust is predicted by
{\texttt{GALFORM}} self-consistently from the cold gas mass and
metallicity, assuming a dust-to-gas ratio which is proportional to the
gas metallicity, while the radius of the diffuse dust component is
assumed equal to that of the star-forming component, whether a
quiescent disk or a burst, and is also predicted by
{\texttt{GALFORM}}.

We show the model predictions in Fig.~\ref{FIRCOlums} separately for
quiescent (left panel) and starburst galaxies (right panel), where
quiescent galaxies are defined as those whose total SFR is dominated
by SF taking place in the galactic disk (i.e. $\rm SFR_{\rm
disk}>SFR_{\rm burst}$). Solid lines show the median of the predicted
IR-CO relation at different redshifts, while errorbars represent the
$10$ and $90$ percentiles of the distributions.  To facilitate the
comparison between quiescent and starburst galaxies, the typical IR
luminosity at a CO luminosity of $10^9$~$h^{-2}\, \rm K\, km\, s^{-1} \,pc^{2}$
is shown as dotted lines in both panels. We see that the separate
$L_{\rm IR}-L'_{\rm CO}$ relations for quiescent and starburst
galaxies depend only slightly on redshift, but that the relation for
starbursts is offset to higher IR luminosities than for quiescent
galaxies. There are two contributions to this offset in the model. The
first is the different SF laws assumed in starbursts and
in galaxy disks. By itself, this results in roughly 40 times
larger IR luminosities at a given $H_2$ mass for starbursts. However,
as already described, we also assume a CO-to-$H_2$ conversion factor
$X$ which is 7 times smaller in starbursts, which causes an offset in
the $L_{\rm IR}-L'_{\rm CO}$ relation in the opposite sense. The
combination of these two effects results in a net offset of roughly a
factor of 6 in $L_{\rm IR}$ at a given $L'_{\rm CO}$.

A similar bimodality in the IR-CO luminosity plane has also been
inferred observationally by \citet{Genzel10} and \citet{Combes11}. However, these results
rely on inferring the CO(1-0) luminosities from higher CO
transitions when the lowest transitions are not available, which 
could be significantly uncertain, as we discuss  
below \citep[see also][]{Ivison11}.

For comparison, we also plot in Fig.~\ref{FIRCOlums} a selection of
observational data. We plot data for local LIRGs and for
UV/optically-selected star-forming galaxies at $z\sim 1-2$ in the left
panel, to compare with the model predictions for quiescent galaxies,
and for local ULIRGs, SMGs at $z\sim 1-3$, and QSOs at $z \sim 0-6$ in
the right panel, to compare with the model predictions for
starbursts. We note that there are important uncertainties in the
observational data plotted for high redshift ($z > 1$) objects. For
these, the IR luminosities are actually inferred from observations at
a single wavelength ($24 \mu$m or $850 \mu$m), using an assumed shape
for the SED of dust emission. (In addition, the \citet{Riechers10}
data on $z>1$ QSOs are actually for FIR, i.e. $40-120 \mu$m,
luminosities, rather than total IR luminosities.) Furthermore, the
CO$(1-0)$ luminosities for $z > 1$ objects are in most cases also not
direct measurements, but are instead inferred from measurements of
higher CO transitions $J \rightarrow J-1$ (usually $4\rightarrow 3$ or
$3\rightarrow 2$). The conversion from $L'_{\rm CO}(J \rightarrow
J-1)$ to $L'_{\rm CO}(1-0)$ is usually done assuming that the
brightness temperature of the CO line is independent of the transition
$J \rightarrow J-1$, as would be the case if the CO lines are emitted
from an optically thick medium in thermal equilibrium at a single
temperature (as appears to be the case in local spiral galaxies). In
this case, the luminosity $L'_{\rm CO}$ is independent of the
transition studied \citep{Solomon05}. However, recent observations
have shown different brightness temperatures for different CO
transitions in some high-redshift galaxies
\citep{Danielson10,Ivison11}. As a result, there could be large errors
in the CO$(1-0)$ luminosities of high-redshift galaxies when they are
inferred from higher transitions.

Comparing the model predictions for the IR-CO relation with the
observational data, we see that the predictions for quiescent galaxies
at $z=0$ are in broad agreement with the observations of nearby LIRGs, 
while at $z=2$ the model predicts partially the location 
of UV/optically-selected star-forming galaxies.
In the case of starburst galaxies, the predicted relation agrees with
the observations of low-redshift ULIRGs and high-redshift SMGs, and
also with the observations of QSOs at both low and high redshift. The
latter is consistent with the suggestion from observations that QSOs
follow the same $L_{\rm IR}$-$L'_{\rm CO}$ relation as starburst
galaxies \citep[e.g.][]{Evans06,Riechers10}. The model is thus able to
explain the $L_{\rm IR}$-$L'_{\rm CO}$ relation for all objects
without needing to include any heating of dust by AGN. { This also 
agrees with previous theoretical predictions which concluded that only higher 
CO transitions are affected by the presence of an 
AGN (e.g. \citealt{Meijerink07}; \citealt{Obreschkow09d}).}

\section{Evolution of scaling relations of the H$_2$ to $\rm HI$ ratio}\label{ScalingRels}

\begin{figure*}
\begin{center}
\includegraphics[trim = 4mm 2mm 3mm 1mm,clip,width=0.99\textwidth]{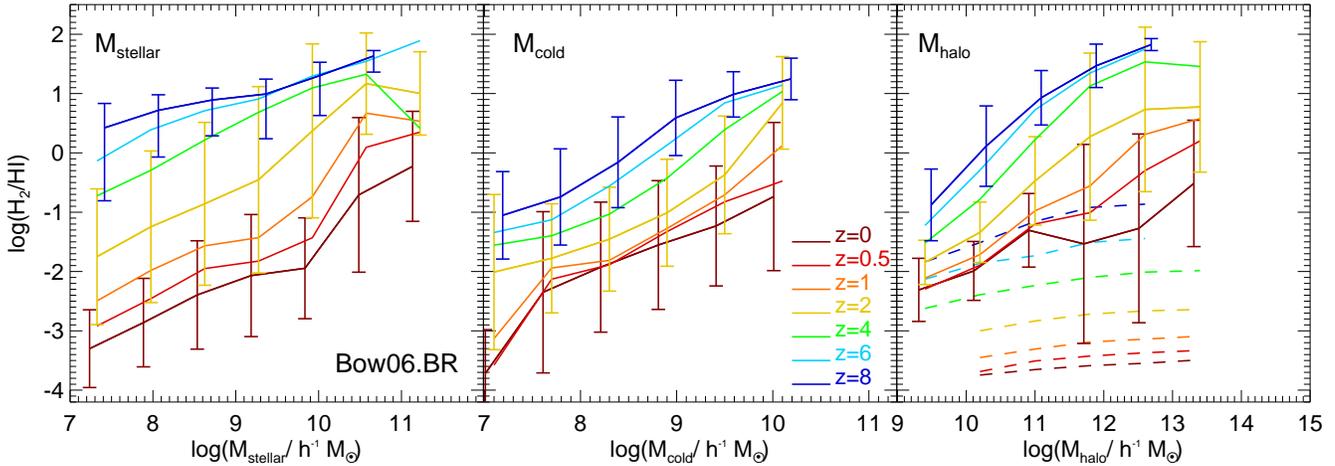}
\caption{Molecular-to-atomic hydrogen ratio as a function of stellar mass (left panel), cold gas mass (middle panel) and DM 
halo mass (right panel) at different redshifts, as labelled, for the 
Bow06.BR model. 
Lines show the medians of the distributions, while 
errorbars show the $10$ to $90$ percentile ranges. For clarity, 
these errorbars are plotted only for three 
different redshifts.
In the right panel, central and satellite galaxies are shown 
separately as solid and dashed lines, respectively. Errorbars are only shown 
for central galaxies in this panel.} 
\label{Scaling2}
\end{center}
\end{figure*}

Our model predicts that 
the HI and H$_2$ MFs are characterised by radically different evolution with redshift. This implies
strong evolution of the H$_2$/HI ratio.
In this section we analyse scaling relations 
of the H$_2$/HI ratio with galaxy properties and track the evolution of these relations towards high
redshift with the aim of understanding what
drives them.

Fig.~\ref{Scaling2} shows the median H$_2$$/$HI ratio as a function of stellar mass, 
cold gas mass and halo mass at different 
redshifts for the Bow06.BR model. Errorbars indicate the $10$ and $90$ percentiles 
of the model distribution in different mass bins. In the right panel, the predictions for 
satellite 
and central galaxies are shown separately as dashed and solid lines, respectively. 
For clarity, percentile ranges are only shown for central galaxies in this panel. 
In the case 
of satellites, the spread  
around the median 
is usually larger than it is for central galaxies.

Fig.~\ref{Scaling2} shows that the H$_2$/HI ratio 
correlates strongly with stellar and cold gas mass 
in an approximately power-law fashion. 
The normalisation of the correlation between H$_2$$/$HI and stellar mass (left panel of Fig.~\ref{Scaling2})  
evolves by two to three orders of magnitude from $z=8$ to $z=0$ 
towards smaller values. The 
evolution is milder (only $\approx 1.5$~dex) if one 
focuses instead on the correlation with cold gas mass. 
Interestingly, the slope of the correlation between the H$_2$$/$HI ratio and 
stellar or cold gas mass hardly 
evolves. {The approximate scalings of the H$_2$/HI mass ratio against stellar and cold gas mass are:}

\begin{eqnarray}
\rm H_2/HI&\approx & 0.01\,\left(\frac{M_{\rm stellar}}{10^{10}{\it h}^{-1} \rm M_{\odot}}\right)^{0.8}\,(1+z)^{3.3},\\ 
\rm H_2/HI&\approx & 0.09\,\left(\frac{M_{\rm cold}}{10^{10}{\it h}^{-1} \rm M_{\odot}}\right)^{0.9} \, (1+z)^{2.4}. 
\label{scalingrels}
\end{eqnarray}
\noindent Note that the relation with $M_{\rm stellar}$ is only valid up to $z=4$. At higher redshifts, 
the slope of 
the correlation becomes shallower. 
The relation with $M_{\rm cold}$, however, has the same slope up to $z=8$. The normalisation 
of the relation of H$_2$$/$HI with $M_{\rm stellar}$ has a stronger dependence on 
redshift compared to the relation with $M_{\rm cold}$.  
However, note that the distribution around the median is quite broad, 
as can be seen from the percentile range plotted in Fig.~\ref{Scaling2}, 
so these expressions have be taken just as an illustration 
of the evolution of the model predictions.

The trend between the H$_2$$/$HI ratio of galaxies and host halo mass 
depends strongly  
on whether central or satellite galaxies are considered. 
In the case of central galaxies 
(solid lines in the right panel of Fig.~\ref{Scaling2}), 
there is a correlation between 
the H$_2$$/$HI ratio and halo mass that becomes shallower 
with decreasing redshift
in intermediate and high mass halos 
($M_{\rm halo}>5\times 10^{11} h^{-1}\rm M_{\odot}$). 
This change in slope is mainly due to the fact that at 
lower redshift ($z\lesssim 1.5$), AGN feedback strongly suppresses SF in central galaxies 
hosted by halos in this mass range, 
so that the stellar mass-halo mass correlation also becomes 
shallower and with an increasing scatter. 
At high redshift, the stellar mass-halo mass correlation 
is tighter and steeper. 
In the case of satellite galaxies, the H$_2$$/$HI ratio does not vary 
with host halo mass, 
but exhibits a characteristic value that depends on redshift. The 
lower the redshift, the lower the characteristic H$_2$$/$HI ratio for satellites. 
Nonetheless, 
the spread around the median, in the case of satellite galaxies,   
is very large, i.e. of $1.5$~dex at $z=0$.
The lack of correlation 
between the H$_2$$/$HI ratio and halo mass and the large scatter are 
due to the dependence of the H$_2$$/$HI ratio on 
galaxy properties (such as stellar and cold gas mass) rather than 
on the host halo mass. 

\begin{figure}
\begin{center}
\includegraphics[trim = 4mm 3.5mm 2mm 2mm,clip,width=0.42\textwidth]{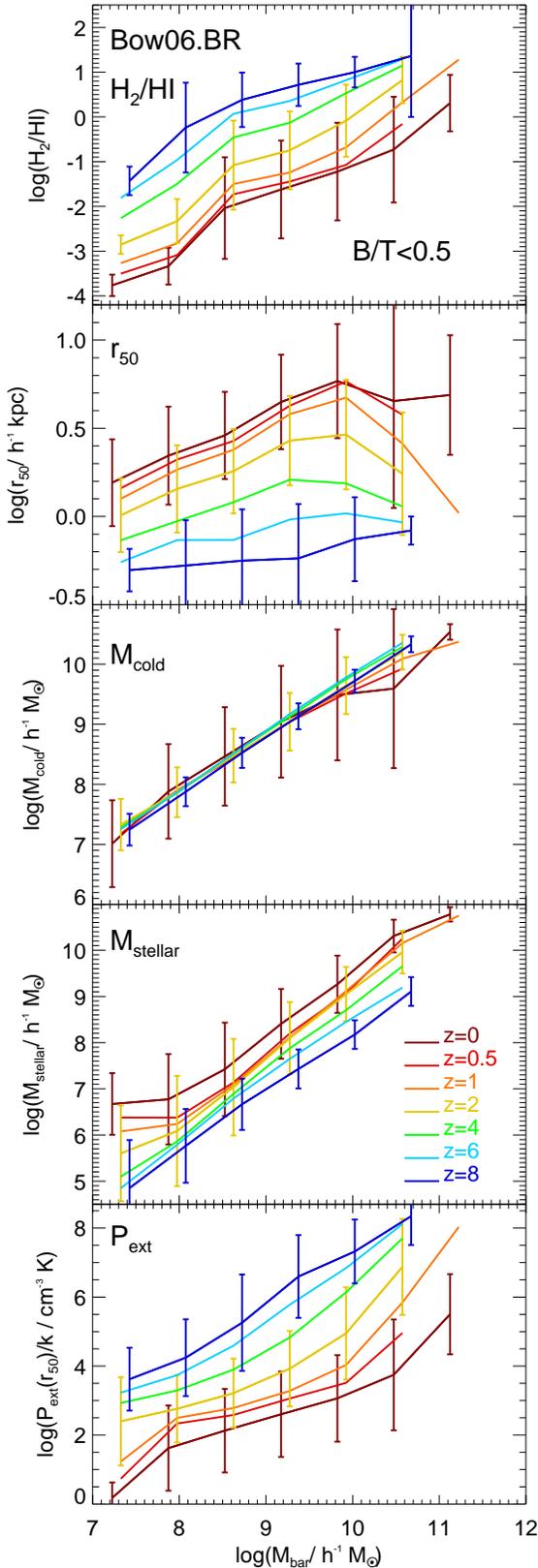}
\caption{Molecular-to-atomic hydrogen ratio (top panel), half-mass radius, $\rm r_{50}$
(second panel), cold gas mass (third panel), 
stellar mass (fourth panel) and midplane hydrostatic pressure at $\rm r_{50}$ (bottom panel), 
as functions of the total cold baryonic mass in the galaxy (i.e. $M_{\rm bar}=M_{\rm
stellar}+M_{\rm cold}$) at different redshifts (as labelled) for the Bow06.BR model. 
Solid lines show the medians of the distributions, while 
errorbars show the $10$ and $90$ percentiles, 
and for clarity are plotted only for three different redshifts.} 
\label{Scaling3}
\end{center}
\end{figure}

In order to disentangle what causes the strong evolution of the H$_2$$/$HI ratio with redshift, we 
study the evolution of the galaxy properties that are directly 
involved in the calculation of the H$_2$$/$HI ratio. 
These are the disk size, cold gas mass and stellar mass, which together 
determine the
hydrostatic pressure of the disk.
Fig.~\ref{Scaling3} shows the H$_2$$/$HI ratio, the 
half-mass radius ($\rm r_{50}$), the 
cold gas mass, the stellar mass and the midplane 
 hydrostatic pressure of the disk at $\rm r_{50}$, as functions 
of the cold baryonic mass of the galaxy, $M_{\rm bar}=
M_{\rm stellar}+M_{\rm cold}$. 
These predictions are for the 
Bow06.BR model. {Note that 
we only plot late-type galaxies (selected as those with a bulge-to-total stellar mass ratio $B/T<0.5$) 
as these dominate the cold gas density at any redshift.}

In general, galaxies display 
strong evolution in size, moving towards larger radii at lower redshifts.
However, this evolution takes place independently 
of galaxy morphology, implying that the galactic size evolution is caused by 
size evolution of the host DM halos (i.e. driven by mergers with other DM halos 
and accretion of dark matter onto halos).
Note that the evolution in radius  
is about an order of magnitude from $z=6$ to $z=0$, which easily explains 
the evolution by two order of magnitudes 
in the H$_2$/HI ratio (which is $\propto P^{0.92}_{\rm ext}\propto \rm r^{-1.84}$). 
By increasing the disk size, the gas density and pressure decrease 
and therefore the molecular fraction decreases. 
The evolution in cold gas mass is mild enough so as not to significantly affect 
the H$_2$/HI ratio. { Observationally, galaxies of the same rest-frame UV luminosity 
seem to be a factor $\approx 2-3$ smaller at $z\approx 6$ than at $z\approx 2$ 
(e.g. \citealt{Bouwens04}; \citealt{Oesch10}), in good agreement 
with the factor of $\approx 2-3$ evolution in size predicted by the Bow06.BR model (see \citealt{Lacey11} 
for a detailed comparison of sizes predicted by the model with the 
observed galaxy population).}

Fig.~\ref{Scaling3} shows that the stellar mass of galaxies, 
at a given baryonic mass,  
increases with decreasing redshift and, therefore, the corresponding 
cold gas mass decreases. This drives the hydrostatic pressure 
to be reduced even further. 
The midplane 
hydrostatic pressure evaluated at $\rm r_{\rm 50}$, $P_{\rm ext}(\rm r_{\rm
50})/k_{\rm B}$ (bottom panel of Fig.~\ref{Scaling3}), evolves 
by more than two orders of magnitude at a given $M_{\rm bar}$ 
over the redshift range plotted. 
Note that the  
typical values of $P_{\rm ext}(\rm r_{\rm 50})/k_{\rm B}$ predicted at $z=0$ are comparable 
to those reported in observations of  
nearby galaxies, $\rm $$P_{\rm ext}/k_{\rm B}=10^3-10^7\,\rm cm^{-3} K$ 
(e.g. \citealt{Blitz06}; \citealt{Leroy08}). 
At higher redshifts, the values of $P_{\rm ext}(\rm r_{\rm 50})$ 
also overlap with the range in which 
the $\Sigma_{\rm H_2}/\Sigma_{\rm HI}$-$P_{\rm ext}$ correlation 
has been constrained observationally at $z=0$.

The evolution in galactic size is therefore the main factor responsible 
for the predicted evolution in the H$_2$$/$HI  
ratio at fixed baryonic mass in the model, with a minor contribution from 
other properties which also contribute to  
determining this quantity (i.e. cold gas and stellar mass). 

\section{Cosmic evolution of the atomic and molecular gas densities}\label{cosmicevo}

\begin{figure}
\begin{center}
\includegraphics[width=0.5\textwidth]{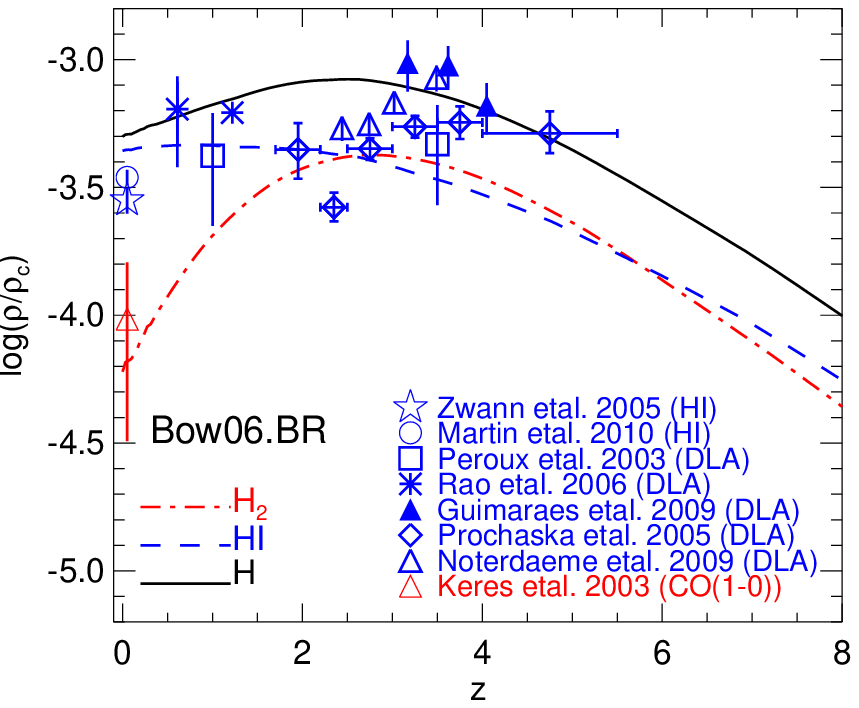}
\includegraphics[width=0.5\textwidth]{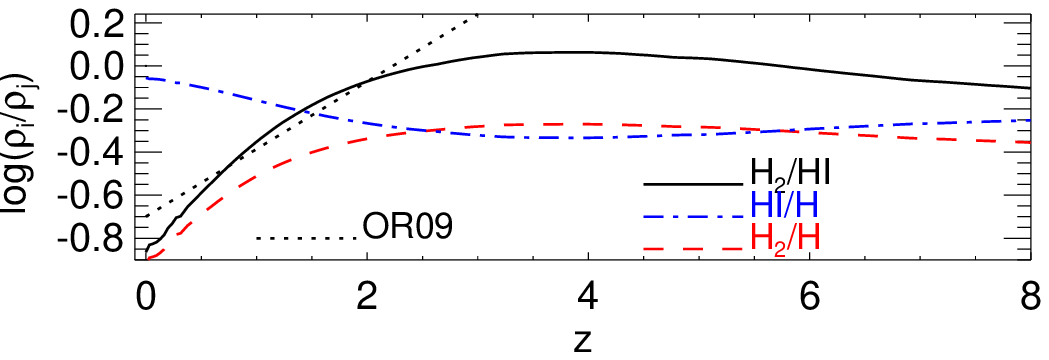}
\caption{{\it Top panel:} Global density of all forms of neutral 
hydrogen gas (solid line), atomic hydrogen (dashed line) 
and molecular hydrogen (dotted-dashed line) in units of the critical density at $z=0$, 
as a function of redshift 
for the Bow06.BR model. Observational estimates of the HI mass density plotted are from \citet{Zwaan05}
and \citet{Martin10} from the HI MF, and \citet{Peroux03}, 
\citet{Prochaska05}, \citet{Rao06}, \citet{Guimaraes09} and
\citet{Noterdaeme09} from DLAs. Also shown
is the local Universe estimate of the H$_2$ mass density from \citet{Keres03}
using the $\rm CO(1-0)$ LF. {\it Bottom panel:} Global atomic-to-total 
neutral hydrogen (dot-dashed line), 
molecular-to-total neutral hydrogen (dashed line) and 
molecular-to-atomic hydrogen (solid line) mass ratios, as functions of 
redshift. For comparison, the evolution of the global molecular-to-atomic
hydrogen ratio predicted by \citet{Obreschkow09c} 
is shown 
as a dotted line.}
\label{denstots}
\end{center}
\end{figure}

The H$_2$/HI ratio depends strongly on galaxy 
properties and redshift. In this section we present
predictions for the evolution of the global density 
of HI and H$_2$.

The top panel of Fig.~\ref{denstots} shows the evolution of the global 
comoving mean density of all forms of hydrogen
(solid line),
HI (dashed line) and H$_2$ (dot-dashed line), in units of the critical density at $z=0$,
$\rho/\rho_{\rm c,z=0}$.
Observational estimates of the HI and H$_2$
mass density at different redshifts and using different techniques are shown using 
symbols (see references in $\S 1$). 
If the reported values of $\rho_{\rm HI}$ and $\rho_{\rm H_2}$ include 
the contribution from helium, then we subtract this when plotting the data.

The model
predicts local universe gas densities in good 
agreement with the observed
ones, which is expected from the good agreement with the HI MF and $
\rm CO(1-0)$ LF. At high redshift, the model predicts $\rho_{\rm HI}$ 
in good agreement with the 
observed density of HI inferred from damped-Ly$\alpha$ systems (DLAs).
Note that in our model, by definition, 
the HI is attached to galaxies (see $\S 2$). 
However, recent simulations by \citet{Faucher-Giguere10}, \citet{Altay10} 
and \citet{Fumagalli11} 
suggest that DLAs might 
not correspond exclusively to HI gas attached to galaxies, 
but also to HI clumps
formed during the rapid accretion of cold gas (e.g.
\citealt{Binney77}; \citealt{Haehnelt98}). 
Thus, the comparison between our 
predictions of $\rho_{\rm HI}$ and the 
values inferred from observations from DLAs 
has to be considered with caution in mind. 

\begin{figure}
\begin{center}
\includegraphics[trim = 2.7mm 2mm 1mm 1mm,clip,width=0.5\textwidth]{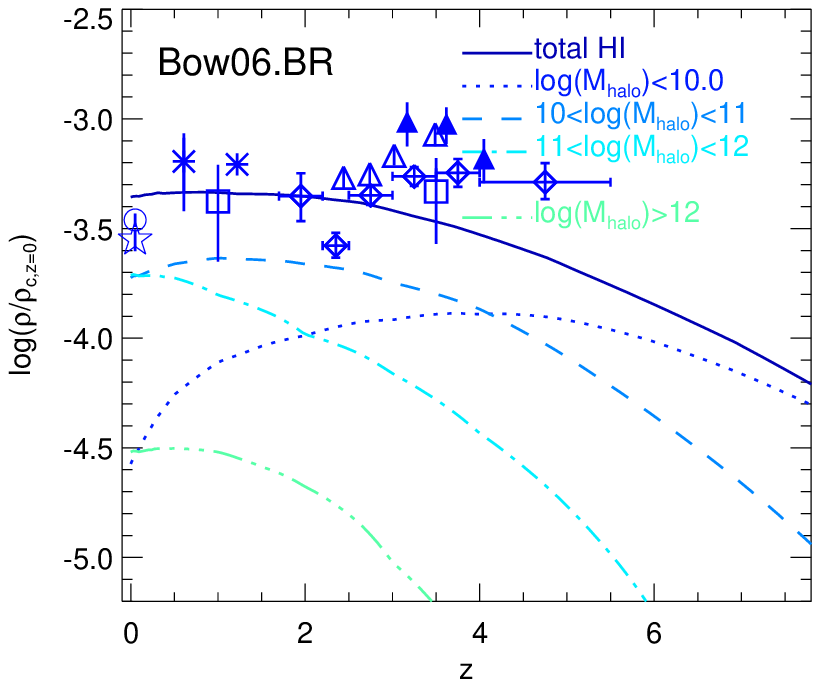}
\includegraphics[trim = 2.7mm 2mm 1mm 1mm,clip,width=0.5\textwidth]{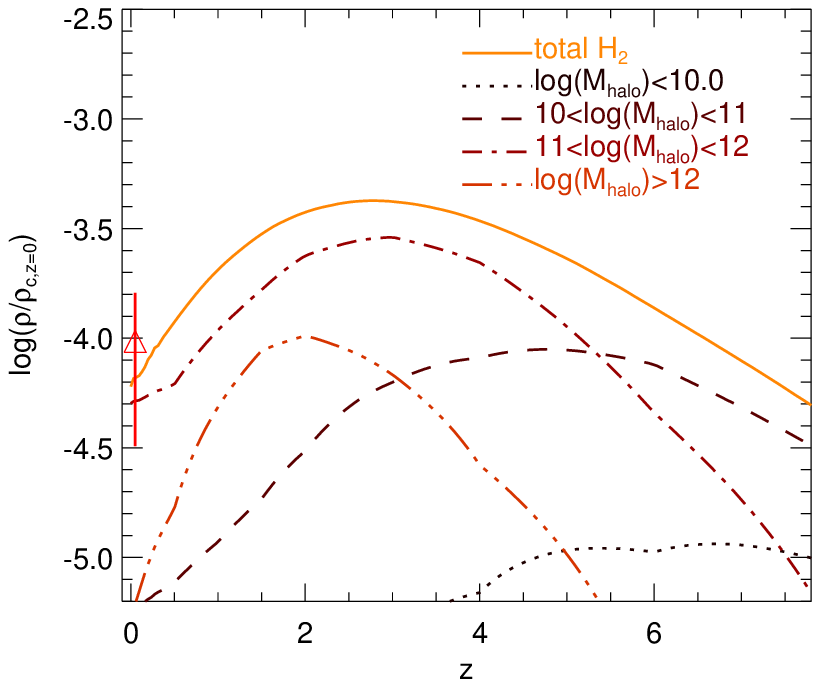}
\caption{Global density of atomic hydrogen (top panel)
and molecular hydrogen (bottom panel) in units of the critical density at $z=0$, as a function of redshift
for the Bow06.BR model. The contributions from different DM halo mass ranges are shown by 
different lines as labelled. 
Observational estimates of the total HI and H$_2$ mass densities are also shown. Symbols correspond 
to different observations as labelled in Fig.~\ref{denstots}.}
\label{denstots2}
\end{center}
\end{figure}

At $z=0$ the HI mass density is much larger 
than that in H$_2$, due to the low molecular fractions 
in extended, 
low pressure galactic disks. At $z\approx 2$, where galactic disks are characterised 
by larger pressure and therefore, higher molecular fractions, 
the density of H$_2$ slightly exceeds that of HI.
This H$_2$ domination extends up to $z\approx 5$, above which HI again becomes  
the principle form of neutral hydrogen. These transitions are clearly seen
in the bottom panel of Fig.~\ref{denstots}, where the evolution of the HI/H,
H$_2$/H and  
H$_2$/HI global ratios are shown. Note that the peak of the H$_2$/HI global ratio is at $z \approx
3.5$. The predicted evolution of the H$_2$/HI global ratio 
differs greatly from the theoretical predictions reported previously 
by Obreschkow \& Rawlings (2009, dotted line in the 
bottom panel of Fig.~\ref{denstots}) and \citet{Power10}, which were obtained from 
postprocessed semi-analytic models (see $\S 2.4$). These authors reported 
a monotonic increase of the H$_2$/HI ratio 
with redshift, even up to global ratios of H$_2$/HI$\approx10$. 
This difference with our results is due in part to resolution effects, since 
these papers used DM halos extracted from 
the Millennium simulation (see $\S 2.1$), and therefore only sample 
haloes with $M_{\rm halo}\ge 
10^{10} \rm h^{-1} M_{\odot}$ at all redshifts.
Consequently, these calculations were 
not able to resolve the galaxies 
that dominate the HI global density  
at high redshift, inferring an artificially high global H$_2$/HI ratio. 
We can confirm this assertion if 
we fix the halo mass resolution in the Bow06.BR model 
at $M_{\rm halo}=10^{10}\, h^{-1} M_{\odot}$, 
whereupon a global ratio of H$_2$/HI$\approx 7$ is attained at
$z=8$. This is in better agreement with the value predicted 
by \citet{Obreschkow09c}, but is still $\sim 1.5$ times lower. 
We interpret this difference as being due to the 
postprocessing applied by Obreschkow et al. rather 
than the self-consistent calculation adopted here.
We have already showed that postprocessing can lead 
to answers which differ substantially from the self 
consistent approach (see Fig~\ref{HIcompall}).  
We can summarise the predicted evolution of the 
H$_2$/HI global ratio, $\rho_{\rm H_2}/\rho_{\rm HI}$, as being 
characterised by three stages,

\begin{eqnarray}
\rm \rho_{\rm H_2}/\rho_{\rm HI}& \approx& 0.13\,(1+z)^{1.7}\quad {\rm for } \quad z\lesssim2\\
& \approx &0.45\,(1+z)^{0.6}\quad {\rm for } \quad 2\lesssim z \lesssim 4\nonumber\\
& \approx &3.7\,(1+z)^{-0.7}\quad {\rm for } \quad z\gtrsim 4.\nonumber
\label{omehah2hi}
\end{eqnarray}

Fig.~\ref{denstots2} shows the contribution to $\rho_{\rm HI}$ (top panel) and $\rho_{\rm H_2}$ 
(bottom panel)  
from halos of different mass. The overall evolution
of $\rho_{\rm HI}$ is always dominated by low and intermediate mass halos with 
$M_{\rm halo}< 10^{12} \,h^{-1} \rm M_{\odot}$. At $z>1$, the 
contribution from halos with masses of 
$10^{11} \,h^{-1} \rm M_{\odot} <M_{\rm halo}< 10^{12} \,h^{-1} \rm M_{\odot}$
 quickly drops, and at $z>4$ the same happens with  
halos in the mass range 
$10^{10} \,h^{-1} \rm M_{\odot} <M_{\rm halo}< 10^{11} \,h^{-1} \rm M_{\odot}$. 
At higher redshifts,  
low mass halos become the primary hosts of HI mass. 
In contrast, the evolution of $\rho_{\rm H_2}$ is always dominated by intermediate and high mass 
halos with $M_{\rm DM}> 10^{11} \,h^{-1} \rm M_{\odot}$, 
supporting our interpretation of the evolution of the HI and H$_2$ MFs. 
This suggests that the weak clustering of HI galaxies reported 
by \citet{Meyer07} and \citet{Basilakos07} is mainly due 
to the fact that HI-selected galaxies are 
preferentially found in low mass halos, while in more massive halos the hydrogen content 
of galaxies has a larger contribution from H$_2$. 

Note that the evolution of 
$\rho_{\rm HI}$ and $\rho_{\rm H_2}$ with redshift is weaker than 
the evolution of the SFR density, $\rho_{\rm SFR}$. As L11 reported,
$\rho_{\rm SFR}$ increases 
by a factor of $\approx 15$ from $z=0$ to $z=3$ 
(the predicted peak of the SF activity).
On the other hand, $\rho_{\rm H_2}$ evolves by a factor of $\approx 7$ from $z=0$ to $z=3$ 
(where the H$_2$ density peaks), while $\rho_{\rm HI}$ evolves weakly, decreasing by a factor only of $1.5$ 
from $z=0$ to $z=5$. The difference in the redshift 
evolution between $\rho_{\rm H_2}$ and 
$\rho_{\rm SFR}$ is due to the contribution from 
starbursts to the latter at 
$z>2$, that drops quickly at lower redshifts. This leads 
to a larger decrease in the global   
$\rho_{\rm SFR}$ compared to $\rho_{\rm H_2}$. { We remind the reader 
that we assume different SF laws for starbursts and quiescent SF, and therefore, 
by construction, different gas depletion timescales (see \S 2.3).}

\begin{figure}
\begin{center}
\includegraphics[trim = 3.3mm 2.5mm 1mm 1mm,clip,width=0.5\textwidth]{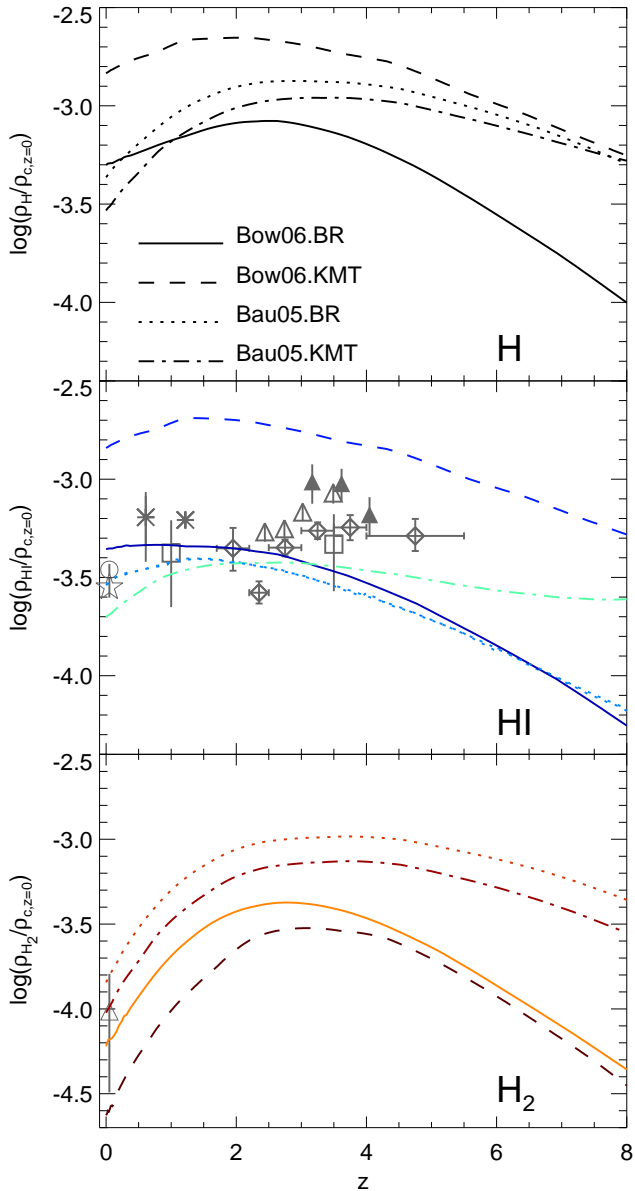}
\caption{Global density of 
all forms of neutral H (top panel), 
HI (middle panel) and H$_2$ (bottom panel) in units 
of the critical density at $z=0$, as 
a function of redshift for the Bow06.BR, Bow06.KMT, Bau05.BR and Bau05.KMT models, 
as labelled in the top panel.
Symbols show observational estimates, as labelled in Fig.~\ref{denstots}.}
\label{GasDensAllmods}
\end{center}
\end{figure}

In order to gain an impression of 
how much the predicted global density evolution can 
change with the modelling of SF, 
Fig.~\ref{GasDensAllmods} shows the global density 
evolution of the neutral H, HI and H$_2$ for the 
Bau05.BR, Bau05.KMT, Bow06.BR and Bow06.KMT models. 
The former three models predict 
very similar global density evolution of HI at $z<4$, while 
at higher redshifts the Bau05.KMT model predicts larger 
global densities mainly 
due to the metallicity dependence 
of the H$_2$/HI ratio in the KMT SF law (see Eq.~\ref{sfreq2}) 
and the typically lower 
gas metallicities in high redshift galaxies. However, 
the Bow06.KMT model 
predicts a global density of HI offset by a factor of $4$ with
respect to  
the observational estimates and the other models 
over the whole redshift range. 
This is caused by the large overprediction 
of the number density of 
intermediate HI mass 
galaxies by this model (Fig.~\ref{HIz0}). 
In the 
case of the global density of H$_2$, all the models 
predict a similar redshift peak at  
$z\approx 3-4$, but with an offset 
in the maximum density attained. 

\section{Summary and conclusions}\label{conclusion} 

We have presented predictions for the HI and H$_2$ content 
of galaxies and their evolution with redshift 
in the context of galaxy formation in the 
$\Lambda$CDM framework. 
We use the extension of the {\texttt{GALFORM}} semi-analytic model 
implemented by Lagos et al. (2010, L11), in which the neutral 
hydrogen in the 
ISM of galaxies is split into its atomic and molecular components. 
To do that we have used either the empirical 
correlation of Blitz \& Roslowsky (2006, BR), between the H$_2$/HI ratio and the 
hydrostatic pressure of the
disk or  
the formalism of Krumholz et al. (2009, KMT) of molecule 
formation on dust grains balanced by UV dissociation.
We assume that the SFR depends directly on 
the H$_2$ content of galaxies, 
as suggested by 
observational results for nearby galaxies  
(e.g. \citealt{Wong02}; \citealt{Solomon05}; \citealt{Bigiel08}; \citealt{Schruba10}). 
We do not change any other model parameter, so as 
to isolate the consequences of changing the SF law on the 
model predictions.

We study the evolution of HI and H$_2$ predicted by 
two models, those of Baugh et al. (2005, Bau05) and 
Bower et al. (2006, Bow06), 
after applying the BR or the KMT SF law prescriptions. 
When applying the BR SF law, we find that the Bau05 and the Bow06 
models predict a present-day HI MF 
in broad agreement with observations by \citet{Zwaan05} and \citet{Martin10},
and a relation between the H$_2$/HI ratio and stellar and 
cold gas mass in 
broad agreement with \citet{Leroy08}. 
When applying the KMT SF law, 
the two models 
fail to predict the observed HI MF at $z=0$. Moreover, L11 
find that the Bau05 model fails to predict the 
observed $K$-band LF at $z>0.5$, regardless of the 
SF law assumed. Hence, we select as our best
model that of 
Bow06 with the BR SF law applied (Bow06.BR). 
This model predicts local Universe estimates 
of the HI and H$_2$ MFs, scaling 
relations of the H$_2$/HI, $M_{\rm H_2}/M_{\rm stellar}$ 
and $M_{\rm HI}/M_{\rm stellar}$ ratios 
with galaxy properties, and also 
the IR-CO luminosity relation 
in reasonable agreement with observations, but also 
gives good agreement with stellar masses, 
optical LFs, gas-to-luminosity ratios, among others (see L11 for a full 
description of the predictions 
of the {\texttt{GALFORM}} model 
when changing the SF law and \citealt{Fanidakis10b} for 
predictions of this model for the AGN population). We show that 
the new SF law applied 
is a key model component to reproduce the local 
HI MF at intermediate and low HI masses 
given that it reduces SF and, therefore, SNe feedback, 
allowing the survival of larger gas reservoirs. 

Using the Bow06.BR model, we can predict 
the evolution of the HI and H$_2$ MFs. We find 
that the evolution of HI is characterised by a 
monotonic increase in the 
number density of massive galaxies with decreasing redshift. 
For H$_2$, the number density of massive galaxies increases from $z=8$ to $z=4$, 
followed by very mild 
evolution down to $z=2$. At $z<2$, the number density 
in this range drops quickly. These radical differences in the 
HI and H$_2$ MF evolution are due to the strong 
evolution in the H$_2$/HI ratio with redshift, due to the dependence on 
galaxy properties. We 
also present predictions for 
the HI mass-stellar mass and HI mass-SFR relations, and 
for the cumulative HI mass density 
of stacked samples of galaxies selected by
stellar mass or SFR at different redshifts, 
which is relevant to the first
observations of HI expected at high redshift. 

At a given redshift, the H$_2$/HI ratio strongly depends on the 
stellar and cold gas masses. The 
normalisation 
of the correlation evolves strongly with redshift, increasing by two orders of magnitude from $z=0$ 
up to $z=8$, in contrast with the very mild evolution in the slope of the correlations. 
We also find that the H$_2$/HI ratio correlates 
with halo mass only for central galaxies, 
where the slope and normalization of the 
correlation depends on redshift. 
For satellite galaxies, no correlation 
with the host 
halo mass is found, which is a result of the 
weak correlation between satellite galaxy  
properties and their host           
halo mass.
We find that the main mechanism driving the evolution 
of the amplitude of the correlation between 
the H$_2$/HI ratio and stellar and cold gas mass, 
and with halo mass in the case of central galaxies, 
is the size evolution of galactic disks. Lower-redshift 
galaxies are roughly an order of magnitude 
larger than $z=8$ galaxies with the same baryonic 
mass (cold gas and stellar mass). 

Finally, we studied the cosmic density evolution
of HI and H$_2$, and find that the former is characterised by 
very mild evolution with redshift, in
agreement with observations of DLAs (e.g. \citealt{Peroux03}; \citealt{Noterdaeme09}),
while the density of H$_2$ increases by a factor of $7$ from $z\approx0$ to $z\approx3$. We also
predict that H$_2$ slightly dominates over the HI 
content of the universe between $z\approx 2-4$ (with a peak 
of $\rho_{\rm H_2}/\rho_{\rm HI}\approx 1.2$ at $z\approx 3.5$), and that it is
preferentially found in intermediate mass DM halos. On the
other hand, we find that the HI is mainly contained in low mass halos. 
The global 
H$_2$/HI density ratio evolution with 
redshift is characterised by a rapid increase from $z=0$ to $z=2$ as 
$\rho_{\rm H_2}/\rho_{\rm HI}\propto \rm (1+z)^{1.7}$, followed by 
a slow increase at $2<z<4$ as $\rho_{\rm H_2}/\rho_{\rm HI}\propto \rm (1+z)^{0.6}$, to finally 
decrease slowly with increasing redshift 
at $z>4$ as $\rho_{\rm H_2}/\rho_{\rm HI}\propto \rm (1+z)^{-0.7}$. 
We conclude that previous studies (\citealt{Obreschkow09c}; \citealt{Power10}) 
overestimated this evolution due in part to halo mass 
resolution effects, as 
the galaxies which dominate the HI density at $z\gtrsim 2$ were not resolved 
in these calculations.

The results of this paper suggest radically 
different evolution of the cosmic densities of HI and H$_2$, as well 
as a strong evolution in the H$_2$/HI ratio in galaxies. 
The next generation 
of radio and submm telescopes (such as ASKAP, MeerKAT, SKA, LMT and ALMA) 
will reveal the neutral gas content of the Universe and 
galaxies up to very high-redshifts,  
and will be able to test the predictions made here.

\section*{Acknowledgements}

We thank Richard Bower, Ian Smail, 
Houjun Mo, Nicolas Tejos, Gabriel Altay, Estelle Bayet and Padeli Papadopoulos 
for useful comments and discussions, and Rachael Livermore and Matthew Bothwell 
for providing the observational data of 
the IR-CO relation and the HI and H$_2$ scaling relations, respectively.
 We thank the anonymous referee for helpful suggestions that improved this work. 
CL gratefully acknowledges a
STFC Gemini studentship. 
AJB acknowledges the support of the Gordon~\&~Betty Moore Foundation.
This work
was supported by a rolling grant from the STFC. 
Part of the calculations for this paper were performed on the ICC 
Cosmology Machine, which is part of the DiRAC Facility jointly funded by STFC, 
the Large Facilities Capital Fund of BIS, and Durham University. 

\nocite{Lah09}
\nocite{Springel05}
\bibliographystyle{mn2e_trunc8}
\bibliography{Mol_Paper}

\label{lastpage}

\end{document}